\documentclass[extra] {gji}

\pdfoutput=1

\usepackage{timet}
\usepackage{times}
\usepackage{graphicx}
\usepackage{lscape}
\usepackage{xcolor}
\usepackage{amsmath}
\usepackage{amssymb}
\usepackage{calrsfs}
\usepackage{bm}

 \usepackage{lineno}

\renewcommand{\cite}{\citet}

\title[Contributions to the geomagnetic secular variation from a reanalysis of core surface dynamics]
	{Contributions to the geomagnetic secular variation from a reanalysis of core surface dynamics}
\author[Barrois et al.]
  {O. Barrois$^1$, N. Gillet$^1$ \& J. Aubert$^2$\\
  $^1$ Univ. Grenoble Alpes, CNRS, ISTerre, CS 40700 F-38058 Grenoble cedex 9, France.\\
  $^2$ Institut de Physique du Globe de Paris, Sorbonne Paris Cit\'e, Univ. Paris Diderot, CNRS, 1 rue Jussieu, F-75005 Paris, France.}
\date{Received XXX; in original form XXX}
\pagerange{\pageref{firstpage}--\pageref{lastpage}}
\volume{XXX}
\pubyear{2017}

\begin{document}

\label{firstpage}

\maketitle

\begin{summary}

We invert for motions at the surface of Earth's core under spatial and temporal constraints that depart from the mathematical smoothings usually employed to ensure spectral convergence of the flow solutions.
Our spatial constraints are derived from geodynamo simulations.
The model is advected in time using stochastic differential equations coherent with the occurrence of geomagnetic jerks.
Together with a Kalman filter, these spatial and temporal constraints enable the estimation of core flows as a function of length and time-scales.
From synthetic experiments, we find it crucial to account for subgrid errors to obtain an unbiased reconstruction.
This is achieved through an augmented state approach. 
We show that a significant contribution from diffusion to the geomagnetic secular variation should be considered even on short periods, because diffusion is dynamically related to the rapidly changing flow below the core surface.
Our method, applied to geophysical observations over the period 1950-2015, gives access to reasonable solutions in terms of misfit to the data.  
We highlight an important signature of diffusion in the Eastern equatorial area, where the eccentric westward gyre reaches low latitudes, in relation with important up/down-wellings.
Our results also confirm that the dipole decay, observed over the past decades, is primarily driven by advection processes.
Our method allows us to provide probability densities for forecasts of the core flow and the secular variation.

\end{summary}


\section{Introduction}

The past decade has seen the advent of geomagnetic data assimilation techniques, aiming at modeling the core state by considering constraints not only from geophysical observations, but also from our knowledge of the core dynamics \citep{fournier2010introduction}. 
This approach, widely developed to study the dynamics of surface envelopes (ocean, atmosphere), is particularly suited if one aims at either predicting or understanding a dynamical systems (this latter activity being usually referred to as reanalysis).
In the context of the geodynamo, reanalyses are promising in the perspective of imaging un-observed quantities (such as the magnetic field, the flow or the buoyancy flux deep in the fluid outer core), and thus isolating mechanisms responsible for the generation of the time-varying Earth's magnetic field. 
On the other hand, forecasts aim at proposing future probability densities for the evolution of the field that constrains our spatial environment, with implication in space weather -- see for instance the damages from cosmic rays on low Earth orbiting satellites as they pass through areas of low magnetic intensity such as the South-Atlantic anomaly \citep{heirtzler2002future,aubert2015geomagnetic}.

Several avenues have been followed to handle those two questions.
One is to use three-dimensional forward simulations of the geodynamo \citep{liu2007observing,fournier2013ensemble} to derive the state of the core (magnetic, velocity and codensity fields) using the primitive induction, momentum and heat equations, given observations of the radial magnetic field at the core--mantle boundary (CMB).
However, because of the huge numerical cost required to reach Earth-like regimes, those simulations are presently run using unrealistic dimensionless parameters, implying too large dissipation processes -- see e.g. the discussions by \cite{Cheng2016} and \cite{bouligand2016frequency}. 
Dynamo simulations are nevertheless able to provide static and kinematic images of the core consistent with geomagnetic field models \citep[e.g.,][]{christensen2010conditions}.
However, their current development prevents from appropriately modeling the dynamics associated with rapid changes of the secular variation (the rate of change of the magnetic field, or SV).

An alternative avenue consists in considering reduced models able to relatively enhance the role played by magnetic forces, as initiated by \cite{canet2009forward} or \cite{labbe2015magnetostrophic} under the quasi-geostrophic (QG) assumption. 
However, such models are not yet operational.
In the absence of entirely satisfying prognostic models, SV predictions propagated by core surface motions have been carried out, using piecewise stationary flows \citep{beggan2009forecasting,whaler2015derivation}. 
These first pragmatic attempts are operational but do not include important information contents, for instance on the temporal correlation of the flow, or the subgrid errors  associated with the unresolved CMB magnetic field at small length-scales.
These issues have been addressed in the framework of stochastic processes \citep[see][]{van2007stochastic}, which appear able to estimate the probability density function (PDF) of time-dependent core surface flows.
\cite{gillet2015planetary} proposed a re-analysis of QG transient motions over 1940--2010 by means of a weak formalism, while \cite{gillet2015stochastic} used instead an Ensemble Kalman Filter \citep[EnKF, see][]{evensen2003ensemble} to predict the magnetic field PDF in the context of the IGRF-12 \citep{thebault2015international}. 
In this latter proof-of-concept study, subgrid errors were accounted for with an augmented state approach \citep[e.g.,][]{reichle2002hydrologic}.
We refer for instance to \cite{miller1999data} for an illustration of the stochastic EnKF efficiency to describe the evolution of the model state PDF, and to \cite{buizza1999stochastic} for the representation of model uncertainties through stochastic processes, and their impact on the prediction scores.

In the present work, we merge for the first time spatial information provided by numerical simulations with temporal constraints brought by specifically chosen stochastic processes. 
The former is obtained by free runs of a three-dimensional geodynamo model, as initiated by \cite{fournier2011}, and  has been previously used to infer series of independent snapshot core flows from geomagnetic field models by \cite{aubert2013flow,aubert2014earth}. 
The latter extends the algorithm developed by \cite{gillet2015stochastic}, in particular by considering a contribution from core surface magnetic diffusion that improves the analysis of \cite{aubert2014earth}. 
Furthermore, here we follow and complement an idea supported by \cite{amit2008accounting}, and derive diffusion from cross-covariances involving up/down-wellings and the gradient of the magnetic field below the CMB.

The present work displays similarities with the work of \cite{baerenzung2014bayesian,baerenzung2016flow}, as it aims to depart from mathematical smoothing often employed to ensure spectral convergence \citep[the large scale hypothesis, see][]{TOG8holme15}, possibly enhancing the footprint of unresolved small length-scale structures in the SV at large length-scales.
Through this work we present the first validation, with synthetic experiments, of the ability to recover time-dependent core flow features. 
It is also the first attempt at multi-epoch assimilation that uses spatial information from geodynamo while analysing recent geomagnetic data.

We present in details our algorithm in section \S\ref{sec: methods}.
In section \S\ref{sec: synthetic} we test and validate our approach with synthetic experiments, in order to quantify our ability to infer information on observable and unobservable quantities of the core state.
Next in section \S\ref{sec: geophys} we apply our algorithm in a geophysical configuration with a re-analysis of the COV-OBS.x1 model \citep{gillet2013stochastic,gillet2015stochastic} over 1950--2015. 
We finally discuss in section \S\ref{sec: discussion} possible applications, such as hypothesis testing or the forecast of the geomagnetic field PDF. 

\section{Models and methods}
\label{sec: methods}
\subsection{Spatial cross-covariances from geodynamo simulations}
\label{sec: stats}

The variables used in the present work are summarized in Table \ref{tab:notations}. 
We use spherical coordinates $(r,\theta,\phi)$, and the associated unit vectors $({\bf 1}_r,{\bf 1}_{\theta},{\bf 1}_{\phi})$.
In the frequency range considered in this study (periods longer than one year), the mantle can be considered as an insulator \citep{jault:2015}.
The potential magnetic field ${\bm B} = - \nabla V$, above the core-mantle boundary (of radius $c=3485$ km), is projected onto spherical harmonics:
\begin{linenomath*}\begin{eqnarray}
\label{eq:HarmoSB}
V(r,\theta,\phi) = a \sum_{n=1}^{n_{b}} \left(\dfrac{a}{r}\right)^{n+1} \sum_{m=0}^{n} \left[g_{n}^{m} \cos(m \phi) + h_{n}^{m} \sin(m \phi)\right]P_n^m(\cos \theta)\,,
\end{eqnarray}\end{linenomath*}
where $\{g_{n}^{m}$, $h_{n}^{m}\}$ are the Gauss coefficients, $a=6371.2$ km is the reference radius of the Earth, and $P_{n}^{m}$ are the Schmidt semi-normalized Legendre functions of degree $n$ and order $m$.
The same decomposition holds for the secular variation $\partial{B}_{r}/\partial t$ with the coefficients $\{\dot{g}_{n}^{m},\dot{h}_{n}^{m}\}$, for which we define the spectrum \citep{lowes74}
\begin{linenomath*}\begin{eqnarray}
\label{eq: spectrum SV}
{\cal R}(n,t)=(n+1)\sum_{m=0}^n \left[{\dot{g}_n^m(t)}^2 +{\dot{h}_n^m(t)}^2 \right]\,,
\end{eqnarray}\end{linenomath*}
and its time average $\left<{\cal R}\right>(n)$. We use the notation 
\begin{linenomath*}\begin{eqnarray}
\label{eq:time average}
\displaystyle \left<X\right>=\frac{1}{t_e-t_s}\int_{t_s}^{t_e} X(t)dt\,,
\end{eqnarray}\end{linenomath*}
 with $[t_s,t_e]$ the studied time-span.
Divergence free surface core motions are expressed as \citep[e.g.][]{bloxham1989simple}
\begin{linenomath*}\begin{eqnarray}
\label{eq:TorPolu}
{\bm u}_H(\theta,\phi) = \nabla \times (T r{\bf 1}_r) + \nabla_{H} (r S)\,,
\end{eqnarray}\end{linenomath*}
with the toroidal $T$ and poloidal $S$ scalars:
\begin{linenomath*}\begin{eqnarray}
\label{eq:HarmoSu}
\left\{
\begin{array}{rl}
\displaystyle
T(\theta,\phi) = &\displaystyle\sum_{n=1}^{n_{u}} \sum_{m=0}^{n} \left[{t_c}_{n}^{m} \cos(m \phi) + {t_s}_{n}^{m} \sin(m \phi)\right]P_n^m(\cos \theta)\\
\displaystyle
S(\theta,\phi) = &\displaystyle\sum_{n=1}^{n_{u}} \sum_{m=0}^{n} \left[{s_c}_{n}^{m} \cos(m \phi) + {s_s}_{n}^{m} \sin(m \phi)\right]P_n^m(\cos \theta)
\end{array}
\right.\,.
\end{eqnarray}\end{linenomath*}
${t_{c,s}}_{n}^{m}$ and ${s_{c,s}}_{n}^{m}$ are the toroidal and poloidal spherical harmonic coefficients, which are stored into a vector ${\bf u}(t)$, of size $N_U=2n_u(n_u+2)$.
Magnetic and velocity fields are truncated at degree respectively $n_{b}$ and $n_{u}$ (see below). 
We define the core flow spatial power spectrum as 
\begin{linenomath*}\begin{eqnarray}
\label{eq: spectra}
{\cal S}(n,t)=\frac{n(n+1)}{2n+1}\sum_{m=0}^n \left[{{t_{c,s}}_n^m(t)}^2  + {{s_{c,s}}_n^m(t)}^2 \right]\,,
\end{eqnarray}\end{linenomath*}
and its time average $\left<{\cal S}\right>(n)$.

To build the spatial prior of our model, we use a forward integration of a geodynamo simulation, the Coupled Earth (CE) model \citep{aubert2013bottom}. 
It solves the momentum, codensity and induction equations under the Boussinesq approximation, for an electrically conducting fluid within a spherical shell (of aspect ratio 0.35 between the inner core and the CMB), assuming no-slip (resp. free-slip) conditions at the inner (resp. outer) boundary. 
It furthermore accounts for a heterogeneous mass-anomaly flux at both the inner and outer boundaries, together with a gravitational coupling between the inner core and the mantle. 
Its construction leads to similarities with the Earth's dynamo from both a static (magnetic field morphology) and a kinematic (secular variation and core flow structure) point of view.

We use $N_{CE}=1505$ realizations from the CE dynamo to infer statistics on the magnetic field and the flow, truncated at respectively $n_u^{CE}=18$ and  $n_b^{CE}=30$. 
All realizations are snapshots of a free run, separated by $90$ years -- 
dimensionless times are scaled into years as in \cite{aubert2015geomagnetic}, following \cite{lhuillier11}.
The dimensionless magnetic field is scaled into physical units by matching its spatial spectrum at the CMB to that of the COV-OBS field model, averaged over 1840--2010.    

We write ${\bf B}$ the vector containing magnetic field coefficients;
for reasons detailed below, we dissociate its resolved component at degrees $n\in[1,n_{o}=14]$, stored in ${\bf b}$, and its unresolved component at degrees $n\in[n_{o}+1,n_b^{CE}]$, stored in a vector $\tilde{{\bf b}}$.
SV coefficients up to degree $\dot{n}_o$ are stored in a vector $\dot{\bf b}$, of size $N_{SV}=\dot{n}_o(\dot{n}_o+2)$. 
We use the notation
\begin{linenomath*}\begin{eqnarray}
\label{eq:ens ave}
\hat{X} = \dfrac{1}{N_{CE}} \sum_{k=1}^{N_{CE}} X^{k}
\end{eqnarray}\end{linenomath*}
to define the ensemble average (or background) over realisations $\{X^k\}_{k\in[1,N_{CE}]}$, which approximates the statistical expectation $\mathbb{E}[X]$.
Cross-covariances between flow coefficients are accounted for through the covariance matrix
\begin{linenomath*}\begin{eqnarray}
\label{eq:Pu}
\textsf{P}_{uu} &=& \mathbb{E}\left[\left({\bf u} - \hat{\bf u}\right)\left({\bf u} - \hat{\bf u}\right)^T\right] \\
&=& \dfrac{1}{N_{CE}-1} \sum_{k=1}^{N_{CE}} 
\left({\bf u}^{k} - \hat{\bf u}\right)
\left({\bf u}^{k} - \hat{\bf u}\right)^{T}\,, \nonumber
\end{eqnarray}\end{linenomath*}
with a similar expression for 
$\textsf{P}_{bb}=\mathbb{E}\left[\left({\bf b} - \hat{\bf b}\right)\left({\bf b} - \hat{\bf b}\right)^T\right]$,  
$\textsf{P}_{\dot{b}\dot{b}}=\mathbb{E}\left[\left(\dot{\bf b} - \hat{\dot{\bf b}}\right)\left(\dot{\bf b} - \hat{\dot{\bf b}}\right)^T\right]$,  
$\textsf{P}_{bu}=\mathbb{E}\left[\left({\bf b} - \hat{\bf b}\right)\left({\bf u} - \hat{\bf u}\right)^T\right]$ and $\textsf{P}_{ub}=\textsf{P}_{bu}^T$.

\begin{table*}
\newlength{\digitwidth} \settowidth{\digitwidth}{\rm 0}
\catcode`?=\active \def?{\kern\digitwidth}
\caption{Summary of the notations used throughout this study.
The state ${\bf x}$ shall here be considered as a generic notation (either ${\bf b}$ or ${\bf u}$ or ${\bf e}$). }
\label{tab:notations}
\centering
\begin{tabular*}{0.91\textwidth}{@{}c@{\extracolsep{\fill}}crr}
\hline
					physical space & spectral space & Meaning  & truncation degree \\
\hline
$B_{r}$			& ${\bf B}$ 			&	radial magnetic field 			& $1-n_b^{CE}$\\
$\overline{B}_{r}$	& ${\bf b}$			&	large-scale radial magnetic field 	& $1-n_{o}$ \\
$\tilde{B}_{r}$ 	& $\tilde{\bf b}$		& 	small-scale radial magnetic field	& $n_{o}+1-n_b^{CE}$ \\
$\bm{u}$		& ${\bf u}$			&	surface core flow				& $1-n_u$\\
$e_r$			& ${\bf e}$			&	subgrid errors					& $1-n_{o}$ \\
$d$				& ${\bf d}$			&	diffusion						& $1-n_{o}$ \\
\hline
				& ${\bf e}^o$			&	main field observation errors		& $1-n_{o}$ \\
				& ${\bf b}^o$			&	main field observations  			& $1-n_{o}$ \\
				& $\dot{\bf e}^o$		&	SV observation errors			& $1-\dot{n}_{o}$ \\
				& $\dot{\bf b}^o$		&	SV observations  				& $1-\dot{n}_{o}$ \\
\hline
				& $\hat{\bf x}(t)$		&	background state 				& \\
				& $\left<{\bf x}\right>$	&	time average state 				& \\
				& ${\bf x}^*(t)$		&	reference state 				& \\
				& ${\bf x}^{\dag}(f)$	&	state time Fourier transform		& \\
				& ${\bf x}^f(t)$		&	forecast		 				& \\
				& ${\bf x}^a(t)$		&	analysis						& \\
				& ${\bf x}'(t)$		&	analysis error ${\bf x}^a-{\bf x}^*$& \\
\hline
				& $n_b^{CE}$			&	truncation degree of CE magnetic field		& 30 \\
				& $n_u$				&	truncation degree of core flow 				& 18 \\
				& $n_{o}$			&	truncation degree of observed magnetic field	& 14 \\
				& $\dot{n}_{o}$		&	truncation degree of observed SV 			& 14 \\
\hline
\end{tabular*}
\end{table*}
 
\subsection{A time-dependent stochastic model}

The evolution of the magnetic field $\bm{B}$ within the Earth's core is governed by the induction equation
\begin{linenomath*}\begin{eqnarray}
\label{eq:full_induction}
\frac{\partial \bm{B}}{\partial t} = \nabla \times (\bm{u} \times \bm{B}) + \eta \nabla^{2} \bm{B}\,,
\end{eqnarray}\end{linenomath*}
where $\eta$ is the magnetic diffusivity.
Contrary to the core flows $\bm{u}$ for which we do not have any direct measurements, the magnetic field at the CMB is estimated via the downward continuation, through an insulating mantle, of records at and above the surface of the Earth. 
Only its radial component $B_r$ is continuous through the CMB. 
Its evolution at the core surface is governed by \citep{TOG8holme15}
\begin{linenomath*}\begin{eqnarray}
\label{eq:induction}
\frac{\partial B_{r}}{\partial t} = -\nabla_{H} \cdot (\bm{u}_H B_{r}) + \eta \nabla^{2} B_{r}\,.
\end{eqnarray}\end{linenomath*}
However, we cannot have a complete access to all terms in the above equation. 
First, the diffusion term in (\ref{eq:full_induction}) can only be partially obtained knowing only $B_r$ at the CMB \citep[see][]{gubbins96}. 
We can nevertheless improve our estimate of diffusion using correlations between the surface field and flow with the magnetic field underneath. 
In practice we do not resolve diffusion by means of a dynamical model. 
It results instead from a linear estimate involving covariance matrices between the core surface flow and the magnetic field at and below the CMB, i.e. diffusion is approximated as $\eta\nabla^2B_r=d({\bf u},B_r)$, where $d$ is a linear operator. 
This point is detailed in \S\ref{sec: diffusion}.

Furthermore, because of the geometric attenuation from the CMB upward to the Earth's surface, and the larger power contained into the lithospheric field at short wave-lengths, the main field is resolved only for degrees $n \le n_{o}=14$.
Only the large-scale fraction of the radial magnetic field $\overline{B}_{r}$ is available in equation (\ref{eq:induction}) to retrieve information on $\bm{u}$.
The unresolved component $\tilde{B}_{r}={B}_{r}-\overline{B}_{r}$ nevertheless generates observable SV: the subgrid electro-motive force (e.m.f.) associated with the unresolved field is a major source of uncertainty in (\ref{eq:induction}), and the principal limitation in the estimation of core motions from geomagnetic data \citep{eymin2005core,pais2008quasi}.
Properly accounting for these subgrid errors is crucial to obtain an unbiased estimate of the core state and its associated posterior errors \citep{gillet2015planetary,baerenzung2016flow}.

In that framework, we shall consider the projection of equation (\ref{eq:induction}) onto large length-scales,
\begin{linenomath*}\begin{eqnarray}
\label{eq:Erind}
\frac{\partial \overline{B}_{r}}{\partial t} = - \overline{\nabla_{H} \cdot (\bm{u}_H \overline{B}_{r})} + e_r + d({\bm u}_H,B_r)\,,
\end{eqnarray}\end{linenomath*}	
with $e_r = - \overline{\nabla_{H}\cdot(\bm{u}_H \tilde{B}_{r})}$ the subgrid errors. 
Just as $B_r$ and $\partial{B}_r/\partial t$ in \S\ref{sec: stats}, $e_r$ and $d$ are expanded into spherical harmonics, stored at each epoch $t$ into vectors ${\bf e}$ and ${\bf d}$.
Hereafter, the e.m.f. term on the r.h.s. of equation (\ref{eq:induction}) is written in matrix form ${\sf A}({\bf B}){\bf u}$, with ${\sf A}$ a matrix of size $N_{SV}\times N_U$.
In equation (\ref{eq:Erind}), the e.m.f. arising from the resolved and unresolved magnetic fields write respectively ${\sf A}({\bf b}){\bf u}$ and ${\bf e}={\sf A}(\tilde{{\bf b}}){\bf u}$.
From realizations $\{\tilde{\bf b}^k,{\bf u}^{k}\}_{k\in[1,N_{CE}]}$ of the CE dynamo we obtain a set of realizations $\{{\bf e}^k\}_{k\in[1,N_{CE}]}$, from which we derive the cross-covariance matrix
$\textsf{P}_{ee}=\mathbb{E}\left[\left({\bf e} - \hat{\bf e}\right)\left({\bf e} - \hat{\bf e}\right)^T\right]$ using an expression similar to (\ref{eq:Pu}).
Note that we consider below $\hat{\bf e}={\bf 0}$ (subgrid errors are a priori unbiased), since from realizations of the CE dynamo we find that the ensemble average of subgrid errors is much less than its associated standard deviation.
  
The evolution of the quantities $\bm{u}_H$ and $e_r$ is now required to advect the large-scale part $\overline{B}_{r}$ of the geomagnetic field.
We consider them as random variables, and model their evolution by means of stochastic differential equations \citep[e.g.,][]{yaglom04}.
The flow is governed by an Auto-Regressive process of order 1 (AR-1), expressed with the formulation
\begin{linenomath*}\begin{eqnarray}
\label{eq:AR1 u}
\frac{{\mathrm d}{\bf u}}{{\mathrm d}t}+\frac{1}{\tau_u}\left({\bf u}-\hat{\bf u}\right)={\bm \zeta}_u(t)\,,
\end{eqnarray}\end{linenomath*}
with ${\bm \zeta}_u$ a white noise process (actually the differential of a Wiener process). 
This choice is guided by the occurrence of geomagnetic jerks at inter-annual to decadal periods, which calls for continuous but not differentiable samples \citep{gillet2015planetary}.
A process such as that described by equation (\ref{eq:AR1 u}) is characterized by a Laplacian correlation function $\exp(-\tau/\tau_u)$, where $\tau$ is the time lag.
For the sake of simplicity, a single (i.e. constant) $\tau_u$ is considered for all flow coefficients; the choice for the value of $\tau_u$ is provided in \S\ref{sec: true state}. 
To ensure that cross-covariances of the time-integrated flow ${\bf u}$ are coherent with ${\sf P}_{uu}$, the random forcing ${\bm \zeta}_u$ is generated at each time-step from the Choleski decomposition of  ${\sf P}_{uu}={\sf U}_u{\sf U}_u^T$ as 
${\bm \zeta}_u=\sqrt{2/\tau_u}{\sf U}_u{\bf w}$, with $\bf w={\cal N}({\bf 0},{\bf 1})$ a normal random vector of unit variance \citep[see][]{gillet2015stochastic}.
The numerical integration of equation (\ref{eq:AR1 u}) is then performed with an Euler-Maruyama scheme \citep{kloeden1992numerical},
\begin{linenomath*}\begin{eqnarray}
\label{eq:ARud}
{\bf u} (t+\Delta t) = {\bf u} (t) - \frac{\Delta t}{\tau_u}\left({\bf u} (t) - \hat{\bf u}\right) 
+  \sqrt{\Delta t}{\bm\zeta}_u(t)\,,
\end{eqnarray}\end{linenomath*}
using a numerical time step $\Delta t=0.5$ years.

We follow \cite{gillet2015stochastic} and also consider subgrid errors $e_r$ as realizations of an AR-1 process,
\begin{linenomath*}\begin{eqnarray}
\label{eq:AR1 e}
\frac{{\mathrm d}{\bf e}}{{\mathrm d}t}+\frac{{\bf e}}{\tau_e}={\bm \zeta}_e(t)\,,
\end{eqnarray}\end{linenomath*}
with ${\bm \zeta}_e$ a white noise processes. 
The choice for an AR1 model is here motivated by the empirical estimate of the time cross-covariances in \cite{gillet2015planetary}.
Indeed, they show a Laplacian-like shape (see their figure 1), with $\tau_e$ almost independent of the spherical harmonic degree and order. 
Accordingly, we use $\tau_e=10$ years for all coefficients entering the vector ${\bf e}$.
We ensure that cross-covariances of the numerically integrated ${\bf e}(t)$ are coherent with ${\sf P}_{ee}$ by using the Choleski decomposition of ${\sf P}_{ee}={\sf U}_e{\sf U}_e{}^T$ and ${\bm \zeta}_e=\sqrt{2 /\tau_e}{\sf U}_e\cal{N}({\bf 0},{\bf 1})$.
${\bf e}$ is then time-stepped with the scheme
\begin{linenomath*}\begin{eqnarray}
\label{eq:AR1 e discret}
{\bf e} (t+\Delta t) = \left(1- \frac{\Delta t}{\tau_e}\right){\bf e} (t)
+  \sqrt{\Delta t}{\bm\zeta}_r(t)\,.
\end{eqnarray}\end{linenomath*}
Finally, the system of equations (\ref{eq:Erind},\ref{eq:AR1 u},\ref{eq:AR1 e}) is integrated to forecast the trajectory of the Earth's core state vector
\begin{linenomath*}\begin{eqnarray} 
\label{eq:StateVector}
{\bf x} = \left[ {\bf b}^T \, {\bf u}^T \,{\bf e}^T\right]^T\,.
\end{eqnarray} \end{linenomath*}

\subsection{Diffusion from the CE dynamo}
\label{sec: diffusion}

The diffusion term in equation (\ref{eq:Erind}) cannot be obtained only from the radial component of the field at the CMB, since its expression also requires Gauss coefficients on a shell just below the CMB, of radius $c^-=c-\delta$. 
In the spectral domain the Laplacian writes
\begin{linenomath*}\begin{eqnarray}
\label{eq:DeltaBr}
\nabla^2 g_{n}^{m} = 
\dfrac{2}{\delta^{2}} (g_{n}^{m-} - g_{n}^{m})
- \dfrac{2(n+1)}{c} g_{n}^{m} \left(\dfrac{1}{\delta} + \dfrac{1}{c}\right) \\
- \dfrac{n(n+1)}{c^{2}} g_{n}^{m}\,, \nonumber
\end{eqnarray}\end{linenomath*}
with $\delta=2.7033$ km -- this last value being inherited from the numerical grid set-up of the CE dynamo -- and $g_{n}^{m-}$ the scalar coefficients at radius $c^-$.
Given the dimension chosen to scale time in the CE dynamo (see above),
we have $\eta=1.16$ m$^2$/s, within the range of expected values \citep{aubert2013bottom}. 

In practice, we show that the knowledge of the surface field and flow allows us to estimate the diffusion at the CMB through covariance matrices.
To this purpose, we store coefficients $d_n^m=\eta\nabla^2 g_{n}^{m}$ from realizations of the CE dynamo in an ensemble of vectors $\{{\bf d}^k\}_{k\in[1,N_{CE}]}$. 
We calculate with an expression similar to (\ref{eq:Pu}) the covariance matrices
${\sf P}_{db} = \mathbb{E}\left[\left({\bf d} - \hat{\bf d}\right)\left({\bf b} - \hat{\bf b}\right)^T\right]$ and
${\sf P}_{du} = \mathbb{E}\left[\left({\bf d} - \hat{\bf d}\right)\left({\bf u} - \hat{\bf u}\right)^T\right]$. 
Then, knowing the flow ${\bf u}$ and the large-scale field ${\bf b}$ at the top of the core at a given epoch, we look for the best linear unbiased estimate (under a Gaussian distribution hypothesis) of diffusion, which given our knowledge of the above cross-covariance matrices is \citep[e.g.][]{rasmussen06} 
\begin{linenomath*}\begin{eqnarray} 
\label{eq: diff linear}
{\bf d}^a=\hat{\bf d} + 
\left[\begin{array}{rl}
\displaystyle {\sf P}_{db} & \displaystyle {\sf P}_{du}
\end{array}\right]
\left[
\left[\begin{array}{rl}
\displaystyle {\sf P}_{bb} &\displaystyle{\sf P}_{bu}\\
\displaystyle {\sf P}_{ub} &\displaystyle{\sf P}_{uu}
\end{array}\right] + 
\left[\begin{array}{rl}
\displaystyle {\sf R}_{bb} &\displaystyle{\sf 0}\\
\displaystyle {\sf 0} &\displaystyle{\sf R}_{uu}
\end{array}\right]
\right]^{-1} \\
\cdot \left[\begin{array}{l}
{\bf b}-\hat{\bf b}\\
{\bf u}-\hat{\bf u}
\end{array}\right]\,, \nonumber
\end{eqnarray} \end{linenomath*}
where the superscript '$a$' stands for `analysis'.
${\sf R}_{bb}$ and ${\sf R}_{uu}$ are `observation' error matrices on vectors ${\bf b}$ and ${\bf u}$.
Note that $\hat{\bf d}$ is not negligible, in particular the average diffusion of the axial dipole in the CE dynamo is significantly non-zero \citep[see][]{finlay2016gyre}.
The estimate (\ref{eq: diff linear}) differs from that of \cite{aubert2013flow,aubert2014earth} where cross-covariances involving the flow were not considered.

\begin{figure*}
\centering
	\includegraphics[width=0.7\linewidth]{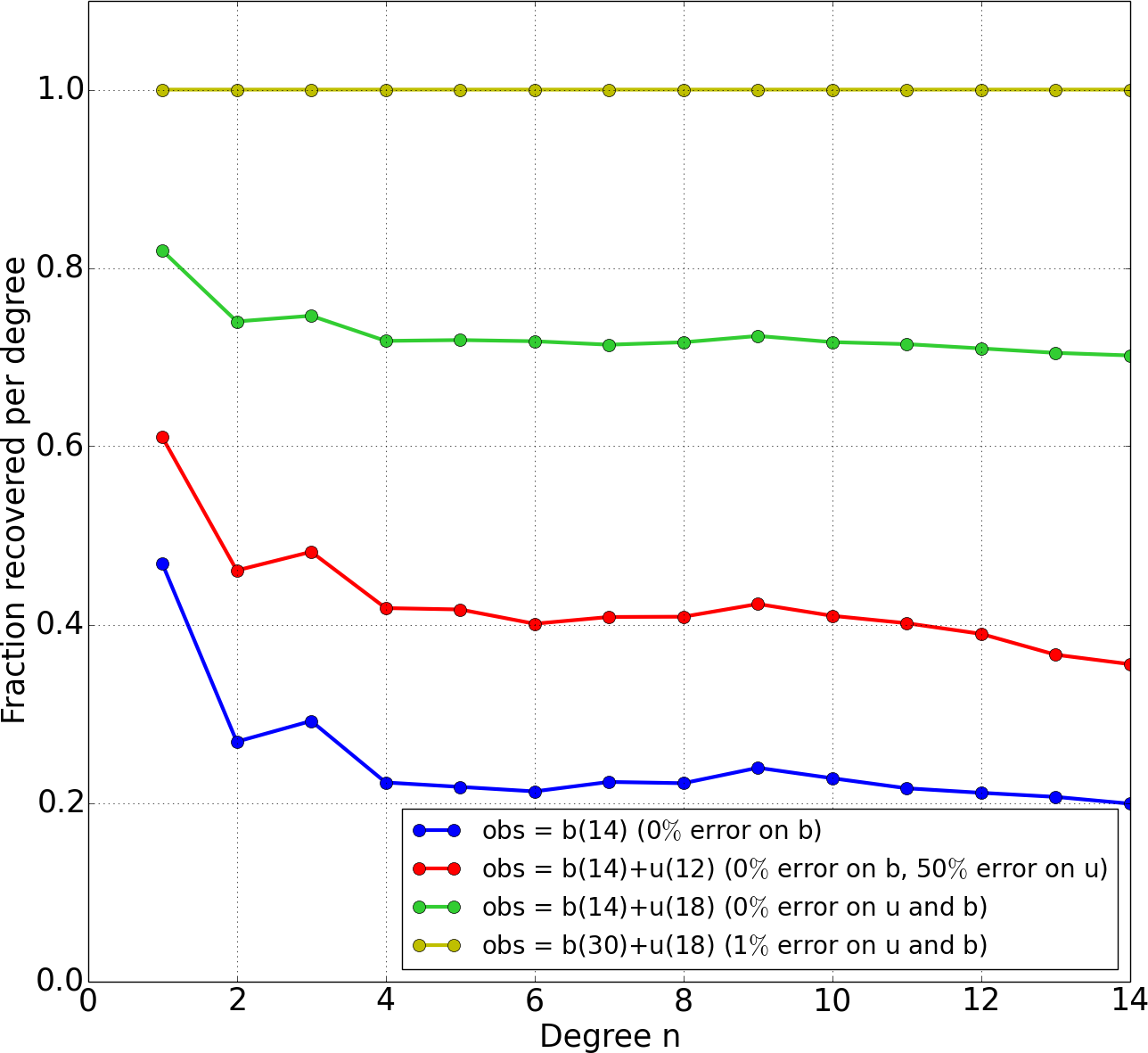}
	\caption{Relative fraction of energy recovered on diffusion as a function of harmonic degree, when estimating diffusion using equation (\ref{eq: diff linear}) where ${\bf b}$ and ${\bf u}$ are snapshots from the CE dynamo, in several configurations:
$B_r$  and ${\bm u}_H$ are almost entirely known up to degrees respectively 30 and 18 (yellow);
$B_r$ and ${\bm u}_H$ are entirely known up to degrees respectively 14 and 18 (green);
$B_r$ is entirely known up to degree 14 and the half of ${\bm u}_H$ (in energy) is known up to degree 12 (red);
$B_r$ only is entirely known up to degree 14 (blue), with no information on ${\bm u}_H$.
}
	\label{fig: diff error}
\end{figure*}

Fig. \ref{fig: diff error} shows how much of the true CE diffusion can be retrieved depending on the information considered in the inverse problem (\ref{eq: diff linear}).
Each curve is obtained from the ratio between the Lowes spectrum of the analysis error (difference between the analysis (\ref{eq: diff linear}) and the CE dynamo diffusion) and the spectrum of the CE dynamo diffusion (spectra are averaged over the $N_{CE}$ snapshots of the CE dynamo).
Ignoring cross-covariances involving the flow, the observable field at degrees below 14 allows us to recover only 20\% of the diffusion from degree 4 onwards (and about 55\% for the lowermost degrees).
In a case where the flow would be entirely known up to degree 18, errors would drop to less than 30\% at high degrees, and to less than 20\% for the dipole.
An intermediate error of about 40 to 60\% is found if 50\% of the flow is known up to degree 12 -- a reasonable error estimate following \cite{gillet2015planetary}.
In the unrealistic case where both the field and the flow are almost entirely known up to degrees respectively 30 and 18, 100\% of the CE diffusion is retrieved from the linear estimate (\ref{eq: diff linear}). 
This shows that, assuming that cross-covariances provided by the dynamo are meaningful, it is possible to retrieve information on the time-changes of surface diffusion from knowledge of only the surface magnetic field and flow.  

These results have important consequences on the analysis of the SV, and encourage us to analyse diffusion in our algorithm (see \S\ref{sec: EnKF}).
Indeed, through equation (\ref{eq: diff linear}) diffusion is now allowed to be responsible for rapid SV changes, because it is linearly related to the flow.
This reflects the modulation by up/down-wellings of magnetic field gradients below the CMB.
Contrary to three-dimensional models that would explicitly calculate diffusion, our two-dimensional model for the advection/diffusion of $B_r$ at the CMB relies on an inversion: 
in the forward integration of equation (\ref{eq:Erind}), the diffusion term $d({\bm u}_H,e_r)$ is obtained from equation (\ref{eq: diff linear}) with ${\sf R}_ {bb}={\sf 0}$ and ${\sf R}_ {uu}={\sf 0}$.
We consider that our mis-estimation of the true diffusion in this case is negligible (cf Fig. \ref{fig: diff error}), in particular in comparison with subgrid errors \citep[and see Fig. 9 in][]{aubert2013flow}.

\subsection{Augmented state Kalman filter}
\label{sec: EnKF}

Now that the forward system is set-up, we describe the algorithm used to invert for the core state.
We seek the most likely trajectory ${\bf x}(t)$ given observations of the main field and its secular variation, statistics on their associated errors, and statistics on the core state described in the above sections.
We write ${\bf b}^o$ and $\dot{\bf b}^o$ the vectors containing observations of the main field and SV Gauss coefficients  (described up to degrees $n_{o}$ and $\dot{n}_{o}$), ${\bf e}^o$ and $\dot{\bf e}^o$ their associated (unbiased) errors, the statistics of which are described by the covariance matrices ${\sf R}_{bb}={\mathbb E}\left[{\bf e}^o{\bf e}^o{}^T\right]$ and  ${\sf R}_{\dot{b}\dot{b}}={\mathbb E}\left[\dot{\bf e}^o\dot{\bf e}^o{}^T\right]$ respectively.

Equations (\ref{eq:Erind},\ref{eq:AR1 u},\ref{eq:AR1 e}) are used to time-step an ensemble of $N_m=50$ realizations of the forecast trajectory  $\{{\bf x}^{kf}(t)\}_{k\in[1,N_m]}$.
We follow \cite{gillet2015stochastic} and use an augmented state approach \citep[see][]{evensen2003ensemble} to invert for  ${\bf x}$. 
Our tool builds on a succession of forecasts and analyses steps, analyses that we perform every $\Delta t^a$ year.
We follow \cite{aubert2014earth} and split the analysis, for each epoch $t_a$ where data are available, in two steps.
First we calculate an ensemble of analyses for ${\bf b}$ from an ensemble of noisy observations ${\bf b}^{ok}$ with the linear filter
\begin{linenomath*}\begin{eqnarray}
\label{eq:EnKF step 1}
\forall k\in [1,N_m],\; \\
{\bf b}^{ka}(t_a)= {\bf b}^{kf}(t_a) + 
{\sf K}_{bb}
\left({\bf b}^{ko}(t_a)-{\bf b}^{kf}(t_a)\right)\,, \nonumber
\end{eqnarray}\end{linenomath*}
with ${\sf K}_{bb}={\sf P}_{bb}^f \left[{\sf P}_{bb}^f + {\sf R}_{bb}\right]^{-1}$ the Kalman gain matrix and ${\sf P}_{bb}^f$ the forecast covariance matrix.

The remaining part of the core state is sought iteratively.
We first obtain an ensemble of diffusion analyses ${\bf d}^{ka}$ using equation (\ref{eq: diff linear}) and an ensemble of ${\bf b}^{ko}$ and flows ${\bf u}^{ka}$.
Next we invert for an ensemble of ${\bf z}^k=\left[{\bf u}^{kT}\; {\bf e}^{kT}\right]^T$ from an ensemble of corrected, noisy observations ${\bf y}^{ko}=\dot{\bf b}^{ko}-{\bf d}^{ka}$ using
\begin{linenomath*}\begin{eqnarray}
\label{eq:EnKF step 2}
\forall k\in [1,N_m],\; \\
{\bf z}^{ka}(t_a)= {\bf z}^{kf}(t_a) + 
{\sf K}_{zz}
\left({\bf y}^{ko}(t_a)-{\sf H}^k{\bf z}^{kf}(t_a) \right)\,, \nonumber
\end{eqnarray}\end{linenomath*}
with ${\sf K}_{zz}={\sf P}_{zz}^f{\sf H}^k{}^T \left[{\sf H}^k{\sf P}_{zz}^f{\sf H}^k{}^T 
+ {\sf R}_{yy}\right]^{-1}$. 
Supposing SV observation errors independent from errors on the diffusion analysis ${\bf d}^a$ (of covariances ${\sf P}_{dd}^a$), one has ${\sf R}_{yy}={\sf R}_{\dot{b}\dot{b}}+{\sf P}_{dd}^a$. 
We discuss below how we approximate the covariance matrices ${\sf P}_{zz}^f$, ${\sf P}_{bb}^f$ and ${\sf P}_{dd}^a$.
The observation operator is ${\sf H}^k=\left[{\sf A}\left({\bf b}^{ka}\right)\; {\sf H}_e \right]$, with ${\sf H}_e$ the identity matrix of rank $N_{SV}$. 
This process (estimation of ${\bf d}$ and ${\bf z}$) is repeated 5 times, which ensures convergence of both the ${\bf z}^{ka}$ and the ${\bf d}^{ka}$.
Note that at the first iteration the diffusion analysis (\ref{eq: diff linear}) is performed with only observations of ${\bf b}^o$ (no contribution from the flow, or ${\sf R}_{uu}$ very large), whereas for the next 4 iterations ${\sf R}_{uu}$ and ${\sf R}_{bb}$ in (\ref{eq: diff linear}) are estimated from the dispersion within the ensemble of solutions.

In contrast with the canonical EnKF \citep{evensen2003ensemble} we do not update, for each analysis step, the forecast covariance matrices 
${\sf P}_{zz}^f=E\left[\left({\bf z}^f-\hat{\bf z}^f\right)\left({\bf z}^f-\hat{\bf z}^f\right)^T\right]$ and 
${\sf P}_{bb}^f=E\left[\left({\bf b}^f-\hat{\bf b}^f\right)\left({\bf b}^f-\hat{\bf b}^f\right)^T\right]$ with the empirical estimate built from the ensemble of realizations. 
Constructing such empirical matrices with well-constrained cross-covariances would indeed require an ensemble of size $N_m$ at least 10 times larger than the size of the matrix to be inverted in equation (\ref{eq:EnKF step 2}) \citep[see][]{fournier2013ensemble}, i.e. in our case several thousands. 
Even if possible (though demanding) to achieve computationally, it is not meaningful to provide such a sophisticated algorithm if we consider that our model does not account for any deterministic dynamics for the flow (see \S\ref{sec: benchmark}). 
Furthermore, any future algorithm including a deterministic physics will most probably be costly, and only operational with ensemble sizes of a few hundreds at most, as it is the case in the community studying surface fluid envelopes \citep[e.g.][]{clayton2013operational}. 
To by-pass this difficulty, numerical approximations are employed, as inflation to avoid ensemble collapse \citep[see][]{hamill2001distance}, or localization to produce well-conditioned matrices \citep[e.g.][]{oke2007impacts}. 
However, this latter is difficult to operate when working in the spectral domain.  

We thus decide to consider frozen matrices in equations (\ref{eq:EnKF step 1}-\ref{eq:EnKF step 2}). 
Let first focus on the analysis for ${\bf z}$. 
We write ${\bf z}^f={\bf z}^a+\delta{\bf z}^f$, with $\delta{\bf z}^f$ the stochastic increment between two analyses. 
Since ${\bf z}^a$ and $\delta{\bf z}^f$ are independent, and $E\left(\delta{\bf z}^f\right)={\bf 0}$, we find 
${\sf P}_{zz}^f={\sf P}_{zz}^a+E\left[\delta{\bf z}^f\delta{\bf z}^{fT}\right]$, with the analysis error covariance matrix ${\sf P}_{zz}^a=E\left[\left({\bf z}^a-\hat{\bf z}^a\right)\left({\bf z}^a-\hat{\bf z}^a\right)^T\right]$. 
The evolution of the PDF for linear AR-1 models such as (\ref{eq:AR1 u}) and (\ref{eq:AR1 e}) can be described analytically \citep[][pp 200-201]{van2007stochastic}:
\begin{linenomath*}\begin{eqnarray} 
\label{eq: scaling}
E\left[\delta{\bf u}^f\delta{\bf u}^{fT}\right]=\alpha_u{\sf P}_{uu}\;\; \mathrm{with} \;\;\alpha_u=2\Delta t^a/\tau_u,  \;\; \\
\mathrm{implying} \;\; {\sf P}_{uu}^f={\sf P}_{uu}^a+\alpha_u{\sf P}_{uu}\,. \nonumber
\end{eqnarray} \end{linenomath*}
A similar expression holds for ${\sf P}_{ee}^f$ where $\alpha_e=2\Delta t^a/\tau_e$. 
The analysis error matrix is in principle ${\sf P}_{zz}^a=\left[{\sf I}-{\sf K}_{zz}{\sf H}\right]{\sf P}_{zz}^f$.
We emphasize in this study two extreme configurations: 
\begin{itemize}
\item[(i)] for a vanishing analysis error (a model state very well constrained by the data) the forecast covariance matrices become ${\sf P}_{uu}^f=\alpha_u{\sf P}_{uu}$ and  ${\sf P}_{ee}^f=\alpha_e{\sf P}_{ee}$;
\item[(ii)] on the opposite, if the innovation vector ${\bf y}^{ko}(t_a)-{\sf H}^k{\bf z}^{kf}(t_a)$ vanishes in (\ref{eq:EnKF step 2}), the forecast covariance matrix shall represent the whole model statistics, which (this is our working hypothesis) are defined by the CE dynamo covariances, i.e. ${\sf P}_{uu}^f={\sf P}_{uu}$ and ${\sf P}_{ee}^f={\sf P}_{ee}$.
\end{itemize}
In both cases cross-covariances between ${\bf u}$ and ${\bf e}$ are ignored when building ${\sf P}_{zz}$.
The latter choice (ii) may appear sub-optimal to recover time changes in the core state, given the temporal correlation of core motions and analyses errors \citep[see Appendix A in][]{gillet2015planetary}, while the former choice (i) might lead to under-estimate the dispersion within the ensemble of flow solutions. 
These issues are discussed further in sections \S\ref{sec: synthetic} and \S\ref{sec: algo}.
Concerning the analysis of ${\bf b}$, the Kalman gain matrix ${\sf K}_{bb}$, in equation (\ref{eq:EnKF step 1}), is almost identity, due to the very small observation error variances entering ${\sf R}_{bb}$ \citep[see][figure 4]{gillet2015stochastic}. 
Thus, the choice for ${\sf P}_{bb}^f$ does not really affect the results: inversions performed with the whole CE dynamo statistics (${\sf P}_{bb}^f={\sf P}_{bb}$) and with its scaled version (${\sf P}_{bb}^f={\Delta t^a}^2{\sf P}_{\dot{b}\dot{b}}$) actually show negligible differences.

The impact of errors on the analysis for diffusion should in principle be considered when building ${\sf R}_{yy}$ for equation (\ref{eq:EnKF step 2}). 
Here again, we will consider two configurations. In the first one ${\sf P}_{dd}^a$ is simply ignored. 
In a second one it is estimated once for all from the statistics of an ensemble of diffusion analysis errors obtained from the CE snapshot realizations using equation (\ref{eq: diff linear}). 
We shall see that these two cases lead to very similar ensemble average solutions, with very close posterior diagnostics (as defined in the next section \S\ref{sec: diagnostics}).  

The present work improves the proof-of-concept study by \cite{gillet2015stochastic}, where diffusion processes were ignored. 
Our scheme makes possible for diffusion errors to alter the forecast, and uncertainties on diffusion analyses will transpire into a larger spread within the ensemble of flow realizations. 
Furthermore, we derive covariances on ${\bm u}_H$ and $e_r$ from the CE dynamo, whereas \citeauthor{gillet2015stochastic} were using a QG topological constraint for the flow, ignoring other spatial cross-covariances in both ${\sf P}_{ee}$ and ${\sf P}_{uu}$ to prevent ill-conditioning.

Our approach also differs from the single epoch algorithm of \cite{aubert2013flow,aubert2014earth}, since the core state is here time-stepped with a (stochastic) dynamical model, carrying information from one epoch to the other.
Our treatment of $e_r$ also differs from that of \citeauthor{aubert2014earth} (see \S\ref{sec: stats}):
we consider that an analysis of $\tilde{\bf b}$ cannot be used to estimate $e_r$ in equation (\ref{eq:Erind}) -- the reason why it is modelled here through the stochastic equation (\ref{eq:AR1 e}).
Indeed, from twin experiments with the CE dynamo, we found that only a small fraction (about 20\%) of the true unresolved field $\tilde{{\bf b}}$ can be recovered from the knowledge of the large scale field ${\bf b}$ and of the cross-covariances between them (not shown).
Note that, in order to tackle this issue, \cite{aubert2015geomagnetic} improved his series of algorithms by generating each analysis within his ensemble of snapshot solutions starting from a random realization sampling the whole CE covariances \citep[and not from the CE average as in][]{aubert2014earth}.
The main steps for the forecast and analysis are summarized in Table \ref{tab:method}.

\begin{table*}
\catcode`?=\active \def?{\kern\digitwidth}
\caption{Summary of the augmented state Kalman Filter as implemented in this study (with an ensemble of size $N_{m} =50$). The core state is defined as ${\bf x}= \left[ {\bf b}^T \, {\bf u}^T \,{\bf e}^T\right]^T$. We refer to the main text for the definition of matrices.}
\label{tab:method}
\centering
\begin{tabular*}{0.71\textwidth}{@{}c@{\extracolsep{\fill}}l}
\hline
1.	&		{\bf Forecast} \\\\
	&		${\mathrm d}{\bf u}/{\mathrm d}t+\tau_{u}^{-1}\left({\bf u}-\hat{\bf u}\right)={\bm \zeta}_u(t)\,,$ \\\\
	&		${\mathrm d}{\bf e}/{\mathrm d}t+\tau_{e}^{-1}{\bf e}={\bm \zeta}_e(t)\,,$ \\\\
	&		${\bf d}({\bf b},{\bf u})=\hat{\bf d} + 
\left[\begin{array}{rl}
\displaystyle {\sf P}_{db} & \displaystyle {\sf P}_{du}
\end{array}\right]
\left[\begin{array}{rl}
\displaystyle {\sf P}_{bb} &\displaystyle{\sf P}_{bu}\\
\displaystyle {\sf P}_{ub} &\displaystyle{\sf P}_{uu}
\end{array}
\right]^{-1}
\left[\begin{array}{l}
{\bf b}-\hat{\bf b}\\
{\bf u}-\hat{\bf u}
\end{array}\right]\,,$ \\\\
	&		${\mathrm d}{\bf b}/{\mathrm d}t = {\sf A}({\bf b}){\bf u} + {\bf e} + {\bf d}({\bf b},{\bf u})\,,$ \\
\hline
2.	&		{\bf Analysis} \\\\
	&		${\bf b}^{a}(t_a)= {\bf b}^{f}(t_a)
+ {\sf P}_{bb} \left[{\sf P}_{bb} + {\sf R}_{bb}\right]^{-1} \left({\bf b}^o(t_a)-{\bf b}^{f}(t_a)\right)\,,$ \\\\
	&		${\bf z}^f(t_a) = \left[{\bf u}^{f}(t_a)^T\,{\bf e}^{f}(t_a)^T\right]^T\,,$ with
		${\sf P}_{zz} = \left[\begin{array}{cc}
\displaystyle \alpha_u{\sf P}_{uu} &\displaystyle{\sf 0}\\
\displaystyle {\sf 0} &\displaystyle\alpha_e{\sf P}_{ee}
\end{array}\right]\,,$ \\\\
	&		${\bf d}^{0} = \hat{\bf d} + {\sf P}_{db}{\sf P}_{bb}^{-1}[{\bf b}^{a}-\hat{\bf b}]\,,$ \\
	&		${\bf y}^{o}=\dot{\bf b}^o-{\bf d}^{0}\,,$ \\\\
	&		for $ i \in [1 : 5]$ \\
	&		\hspace*{1cm} ${\bf z}^{a}(t_a)= {\bf z}^{f}(t_a)
+ {\sf P}_{zz}{\sf H}^T \left[{\sf H}{\sf P}_{zz}{\sf H}^T 
+ {\sf R}_{yy}\right]^{-1}
\left({\bf y}^{o}(t_a)-{\sf H}{\bf z}^{f}(t_a) \right)\,,$ \\
	&		\hspace*{1cm} ${\bf d}^{a}=\hat{\bf d} + 
\left[\begin{array}{rl}
\displaystyle {\sf P}_{db} & \displaystyle {\sf P}_{du}
\end{array}\right]
\left[
\left[\begin{array}{rl}
\displaystyle {\sf P}_{bb} &\displaystyle{\sf P}_{bu}\\
\displaystyle {\sf P}_{ub} &\displaystyle{\sf P}_{uu}
\end{array}\right] + 
\left[\begin{array}{rl}
\displaystyle {\sf R}_{bb} &\displaystyle{\sf 0}\\
\displaystyle {\sf 0} &\displaystyle{\sf R}_{uu}
\end{array}\right]
\right]^{-1}
\cdot \left[\begin{array}{l}
{\bf b}^{a}-\hat{\bf b}\\
{\bf u}^{a}-\hat{\bf u}
\end{array}\right]\,,$ \\
	&		\hspace*{1cm} ${\bf y}^{o}=\dot{\bf b}^o-{\bf d}^{a}\,,$ \\
	&		end \\
\hline
\end{tabular*}
\end{table*}

\subsection{Posterior diagnostics}
\label{sec: diagnostics}

We now define several diagnostics that will be used to evaluate the quality of our algorithm using synthetic experiments (section \S\ref{sec: synthetic}). To do so, we target a reference trajectory ${\bf x}^*$, obtained by numerical integration of the forward model. 
For all three vectors ${\bf v}={\bf u}, {\bf e}$ and ${\bf d}$, we define the bias between the ensemble average and the reference trajectories,
\begin{linenomath*}\begin{eqnarray}
\label{eq: delta_x}
{\bm \delta}_{v}(t)=\displaystyle \hat{\bf v}(t) - {\bf v}^*(t)\,.
\end{eqnarray}\end{linenomath*}
We additionally define the dispersion within the ensemble of state solutions,
\begin{linenomath*}\begin{eqnarray}
\label{eq: epsilon_x}
{\bm \epsilon}_v(t)=\displaystyle\sqrt{
\dfrac{1}{N_m-1}\sum_{k=0}^{N_m} \left[ {\bf v}^k(t) - \hat{\bf v}(t) \right]^2}\,.
\end{eqnarray}\end{linenomath*}
The power spectrum for the flow reference trajectory ${\bm u}^*$, dispersion ${\bm \epsilon}_u$ and bias ${\bm \delta}_u$ are respectively ${\cal S}_{*}(n,t)$, ${\cal S}_{\epsilon}(n,t)$ and ${\cal S}_{\delta}(n,t)$. 
We write ${\cal D}(n,t)$ and ${\cal E}(n,t)$ the Lowes spectra for respectively diffusion and subgrid errors, using an expression similar to that of equation (\ref{eq: spectrum SV}). 
$({\cal D}_*,{\cal E}_*)$ and $({\cal D}_{\delta},{\cal E}_{\delta})$ stand respectively for the spectra of the reference trajectories $({\bf d}^*,{\bf e}^*)$ and of the ensemble average bias $({\bm \delta}_d,{\bm \delta}_e)$. 

From these we calculate several misfits for unobserved quantities (surface core flow, subgrid errors and diffusion) at the analysis steps, normalized to the reference state:
\begin{linenomath*}\begin{eqnarray}
\label{eq: misfit_chi}
\chi_u^2= \dfrac{
\displaystyle\sum_{n=1}^{n_u} \left< {\cal S}_{\delta}^a\right>(n) }{ 
\displaystyle\sum_{n=1}^{n_u} \left< {\cal S}_*\right>(n) }\;\textnormal{, } \;\;
\chi_e^2= \dfrac{
\displaystyle\sum_{n=1}^{n_{\dot o}} \left< {\cal E}_{\delta}^a\right>(n) }{ 
\displaystyle\sum_{n=1}^{n_{\dot o}} \left< {\cal E}_*\right>(n) } \;\\
\textnormal{and}\;\; \chi_d^2= \dfrac{
\displaystyle\sum_{n=1}^{n_{\dot o}} \left< {\cal D}_{\delta}^a\right>(n) }{ 
\displaystyle\sum_{n=1}^{n_{\dot o}} \left< {\cal D}_*\right>(n) }\,. \nonumber
\end{eqnarray}\end{linenomath*}
The superscript `a' for the spectra at numerator means those are calculated for the analysis vectors ${\bm \delta}_v^a=\hat{\bf v}^a - {\bf v}^*$. We recall that brackets stand for time-averaged spectra.  
We also calculate the error with respect to the reference state normalized to the spread within the ensemble \citep[e.g.][]{sanchez2016modelling},
\begin{linenomath*}\begin{eqnarray}
\label{eq: misfit_xi}
\displaystyle \xi_u^2(t) = 
\sum_i
\dfrac{\left( \hat{\sf u}_i(t)-{\sf u}_i^*(t) \right)^2}{N_U{{\epsilon}_u}_i(t)^2}\;\textnormal{, } \;\;
\displaystyle \xi_e^2(t) = 
\sum_i
\dfrac{\left( \hat{\sf e}_i(t)-{\sf e}_i^*(t) \right)^2}{N_{SV}{{\epsilon}_e}_i(t)^2}\; \\
\textnormal{and} \;\; \displaystyle \xi_d^2(t) = 
\sum_i
\dfrac{\left( \hat{\sf d}_i(t)-{\sf d}_i^*(t) \right)^2}{N_{SV}{{\epsilon}_d}_i(t)^2} \nonumber
\,.
\end{eqnarray}\end{linenomath*}
If such quantities are larger (resp. lower) than one, the dispersion within the ensemble under- (resp. over-) estimates the errors to the reference state. 

We shall finally consider the Fourier transform ${\bf u}^{\dag}(f)$ of the time series ${\bf u}(t)$, with $f$ the frequency, from which we build a power spectrum ${\cal S}^{\dag}(n,f)$ with an expression similar to (\ref{eq: spectra}). 
Writing ${\cal S}^{\dag}_*$ the spectrum for the reference trajectory ${\bm u}^*$ and $\hat{\cal S}^{\dag}_{\delta}$ the spectrum for ${\bm \delta}_u^a=\hat{\bm u}^a-{\bm u}^*$, we construct the ratio
\begin{linenomath*}\begin{eqnarray}
\label{eq: chart (f,n)}
\displaystyle {\cal C} (n,f)=\frac{\displaystyle\hat{\cal S}^{\dag}_{\delta}(n,f)}{\displaystyle \hat{\cal S}^{\dag}_*(n,f)}\,. 
\end{eqnarray}\end{linenomath*}
This quantity characterizes our ability to recover core flow time changes: it is zero if the average analysis perfectly matches the reference trajectory, and about unity or greater if the average analysis completely misses the reference trajectory.

\section{Results}
\label{sec: results}

\subsection{Synthetic experiments}
\label{sec: synthetic}

\subsubsection{Construction of the reference trajectory}
\label{sec: true state}

In order to test our algorithm and validate our approach, we first use our method in a synthetic configuration, based on twin experiments. 
This allows us to quantify how much of the core motions can be retrieved and to isolate key ingredients in the inversion scheme. 
In this step before an application to geophysical data, we attempt at building a realistic synthetic model. 
The reference surface core flow ${\bm u}_H^*$ is described up to degree $n_u = 18$ and is numerically integrated using equation (\ref{eq:AR1 u}), using $\tau_u = 30$ years. 
Because our model accounts here for a non-zero background solution, we consider a value of $\tau_u$ shorter than the 100 years preferred by \cite{gillet2015planetary}, but still significantly longer than both $\Delta t^a$ (here equal to 1 yr) and $\tau_e$. 
The reference magnetic field $B_r^*$ is truncated at degree $n_b^{CE} = 30$, and advected with
\begin{linenomath*}\begin{eqnarray}
\label{eq: Br true}
\frac{\partial B_r^*}{\partial t} = - \nabla_{H} \cdot \left({\bm u}_H^* B_{r}^*\right) + d \left({\bm u}_H^*,B_{r}^*\right)\,.
\end{eqnarray}\end{linenomath*}

Diffusion for the reference trajectory should in principle be estimated from equation  (\ref{eq: diff linear}), with ${\sf R}_{ub}$ involving the magnetic field up to $n_b^{CE}$.
However, accounting for cross-covariances with many unresolved field coefficients leads to a ill-conditioned matrix, because of the limited amount of realizations of the CE dynamo.
We thus decide to ignore cross-covariances between the flow ${\bf u}$ and the unresolved magnetic field $\tilde{\bf b}$ -- at degrees $n\in[15,30]$ -- when estimating magnetic diffusion, i.e. diffusion of small length-scales field coefficients is not directly influenced by the flow.
Note that we have tested several intermediate configurations (e.g. gradually smoothing these cross-covariances) with no significant difference on the statistics of the large-scale (observed) magnetic field. 

\begin{figure*}
\centering
	\includegraphics[width=.7\linewidth]{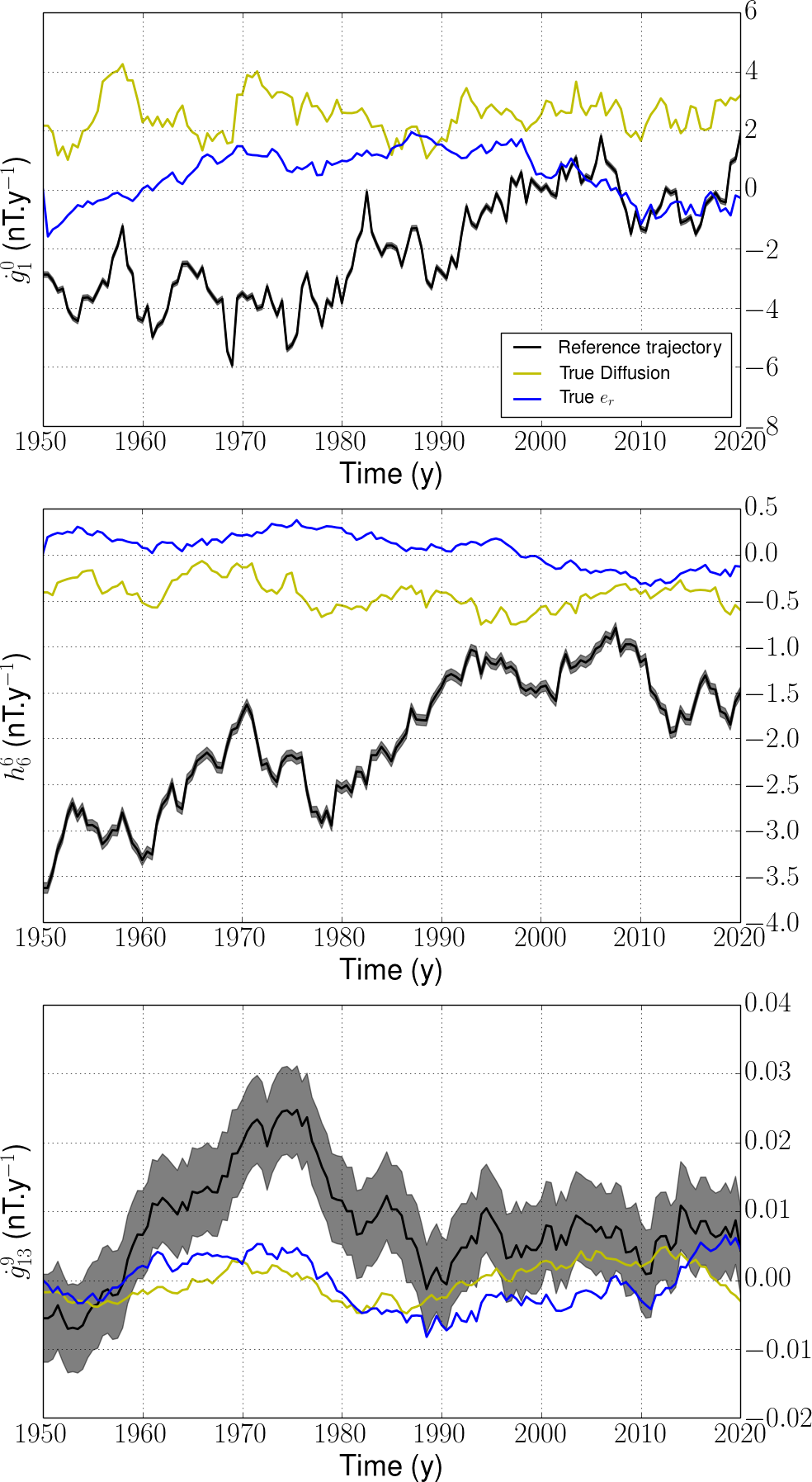}
	\caption{Time series of SV coefficients (black, $\pm \sigma$ the SV observation error in grey shaded area) for the reference trajectory, superimposed with the contributions from diffusion (yellow) and subgrid errors (blue). From top to bottom: $g_1^0, h_6^6$ and $g_{13}^9$.}
	\label{fig:Synt_13_SV series}
\end{figure*}

We initialize the reference trajectory from one realization of the CE dynamo, before the stochastic model defined by equations (\ref{eq:AR1 u},\,\ref{eq: Br true}) is time-stepped from epoch $t_s=1950$ to $t_e=2020$.
Before to go further, we emphasize that our forward model has been constructed such that a non-negligible part of the SV is associated with diffusion (about 10\% of the total SV in power, for all length-scales, see Figure \ref{fig:Synt_13_SV series}). 
Temporal variations of diffusion correlate with those of the flow. 
As a consequence, the contribution from diffusion is not restricted to low frequencies.
At first sight, this may appear surprising since diffusion derives from the slowly varying main field. 
However, radial diffusion at the CMB is enslaved to the magnetic field at and below the core surface, which is dynamically coupled to core motions. 
The link between diffusion and a core flow stretching and twisting magnetic field lines below the CMB transpires in the analysis illustrated with Figure \ref{fig: diff error}. 
As a consequence, diffusion is potentially responsible for rapid SV changes at the CMB, as shown with the reference trajectories of SV Gauss coefficients of different orders in Figure \ref{fig:Synt_13_SV series}.
We have yet to demonstrate that temporal variations of the diffusion are linked to rapid flow variations in a fully self-consistent dynamical model run at parameters closer to Earth's core values.
However, for the mechanistic reasons stated here, we anticipate that this may be the case in the Earth's core, and we have thus constructed our direct model accordingly.

\subsubsection{Re-analysis performances: comparative tests}
\label{sec: reanalysis twin}

We consider below five configurations, with properties summarized in Table \ref{tab:cases}.
We investigate the impact of accounting for subgrid errors and diffusion in the core state, in the case where we do not scale the model cross-covariances (cases A, B and C). 
We further analyse the improvement brought by considering scaled model cross-covariances (case D), with both diffusion and $e_r$ entering the model state. 
These four cases A to D are run while ignoring ${\sf P}_{dd}^a$ when building ${\sf R}_{yy}$. 
A last case E is investigated, where we account for ${\sf P}_{dd}^a$ as described in \S\ref{sec: EnKF} (and otherwise similar to the configuration D). It will be discussed at the end of this section.

\begin{table}
\caption{Posterior diagnostics of equation (\ref{eq: misfit_chi}) for the analysed models ${\bf x}^a(t)$ in the four cases investigated in the synthetic reanalysis. 
The last column corresponds to the flow misfit, but considering the velocity field only up to degree $n=8$.}
\label{tab:cases}
\centering
\begin{tabular*}{0.97\columnwidth}{@{}r@{\extracolsep{\fill}}rrrrcccc}
\hline
Case	& $d$ & $e_r$  & scaled ${\sf P}$  & ${\sf P}_{dd}^a$ & $\chi^2_d$ & $\chi^2_e$ & $\chi^2_{u}$ & $\chi^2_{u{\sf [n\le8]}}$ \\
\hline
A		& yes	 & yes	&	no	&  no   &	0.59		&	1.59		&	0.55		&	0.31 \\
B		& yes	 & no	&	no	&  no   &	1.76		& $\diagup$	&	1.51		&	0.70 \\
C		& no		 & yes	&	no	&  no   & $\diagup$  &	1.73		&	0.62		&	0.33 \\
D		& yes	 & yes	&	yes	&  no   &	0.59		&	1.75		&	0.54		&	0.30 \\
E		& yes	 & yes	&	yes	&  yes  &	0.59		&	1.68		&	0.53		&	0.30 \\
\hline
\end{tabular*}
\end{table}

We initialize the flow and the field from a random draw within the CE realizations, before we perform the re-analysis of the core state with the algorithm presented in \S\ref{sec: EnKF}. 
Data error statistics entering ${\sf R}_{bb}$ and ${\sf R}_{\dot{b}\dot{b}}$ are estimated as the COV-OBS.x1 uncertainties \citep{gillet2015stochastic} evaluated in 2010 (during the satellite era), ignoring cross-covariances.
Together with the reference model trajectory $B_r^{*}$, these statistics are used to build an ensemble of $N_m = 50$ realizations of noisy Gauss coefficient observations,
\begin{linenomath*}\begin{eqnarray}
\label{eq: noisy data}
\displaystyle \forall k\in[1,N_m], \\
{\bf b}^{ok}(t) = {\bf b}^{*}(t) + {\bf e}^{ok}(t) 
\;\mathrm{with}\; E({\bf e}^{ok}{\bf e}^{okT})={\sf R}_{bb}\,. \nonumber
\end{eqnarray}\end{linenomath*}
We use an equivalent process to build an ensemble of $\dot{\bf b}^{ok}$ from $\dot{\bf b}^{*}$ and ${\sf R}_{\dot{b}\dot{b}}$. The SV observation error spectrum is shown in Figure \ref{fig:Synt_13_SV}. 

\begin{figure*}
\centering
	\includegraphics[width=1.0\linewidth]{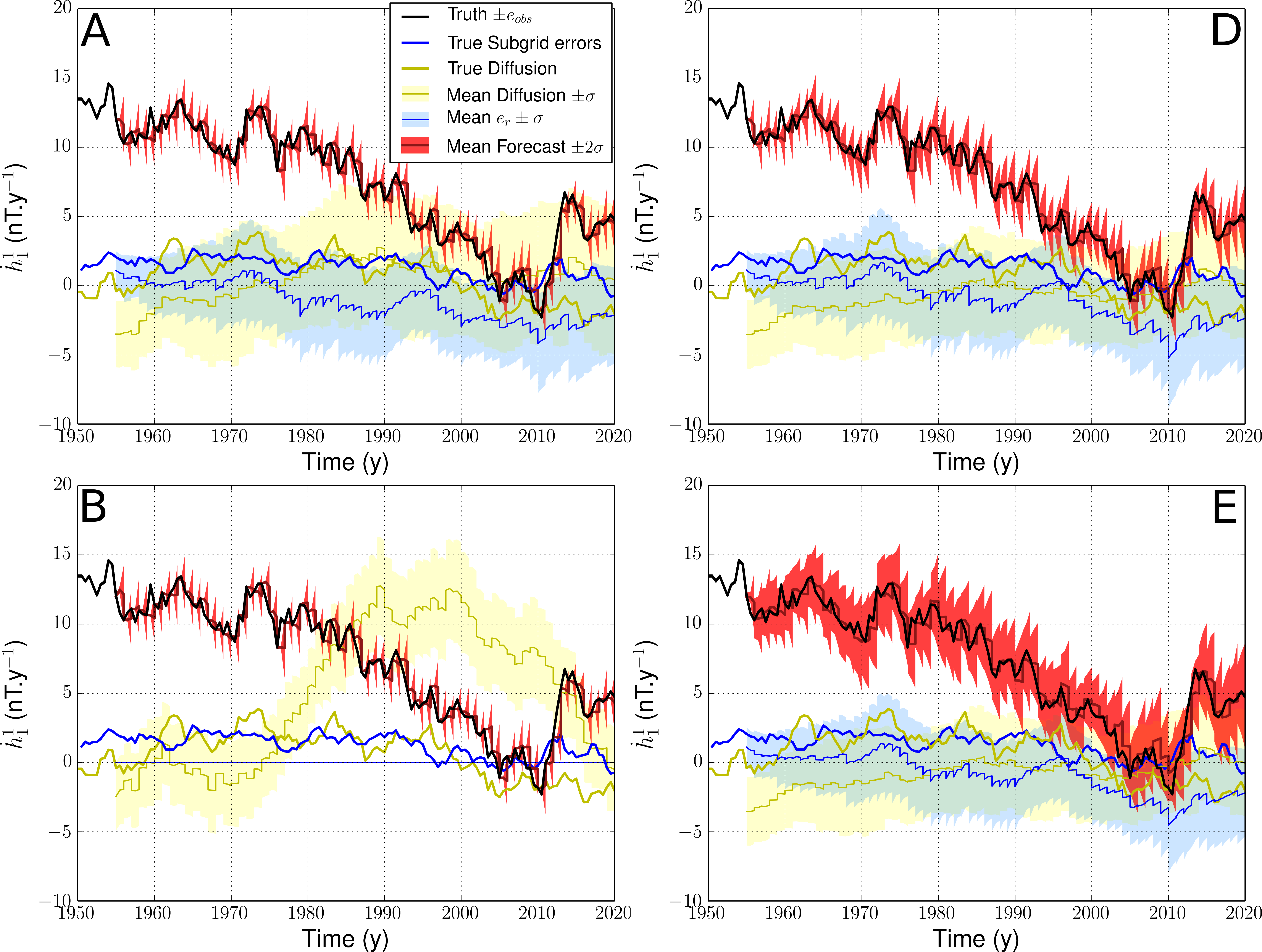}
	\caption{Series of SV coefficients $\dot{h}_{1}^{1}$ for synthetic experiments: 
comparison of our model predictions from $N_m = 50$ reanalyses (average in dark red, $\pm 2 \sigma$ in light red) with the synthetic observations (reference trajectory in black, $\pm\sigma$ observation errors in grey). 
Contributions from $e_r$ (average analysis in blue, $\pm\sigma$ in light blue) and from diffusion (average analysis in yellow, $\pm\sigma$ in light yellow) are also shown, with the reference diffusion in thick yellow and the reference subgrid errors in thick blue.
The four cases A (top left), B (bottom left), D (top right) and E (bottom right) are shown.}
	\label{fig:Synt_13_dg3}
\end{figure*}

\begin{figure*}
\centering
	\includegraphics[width=.7\linewidth]{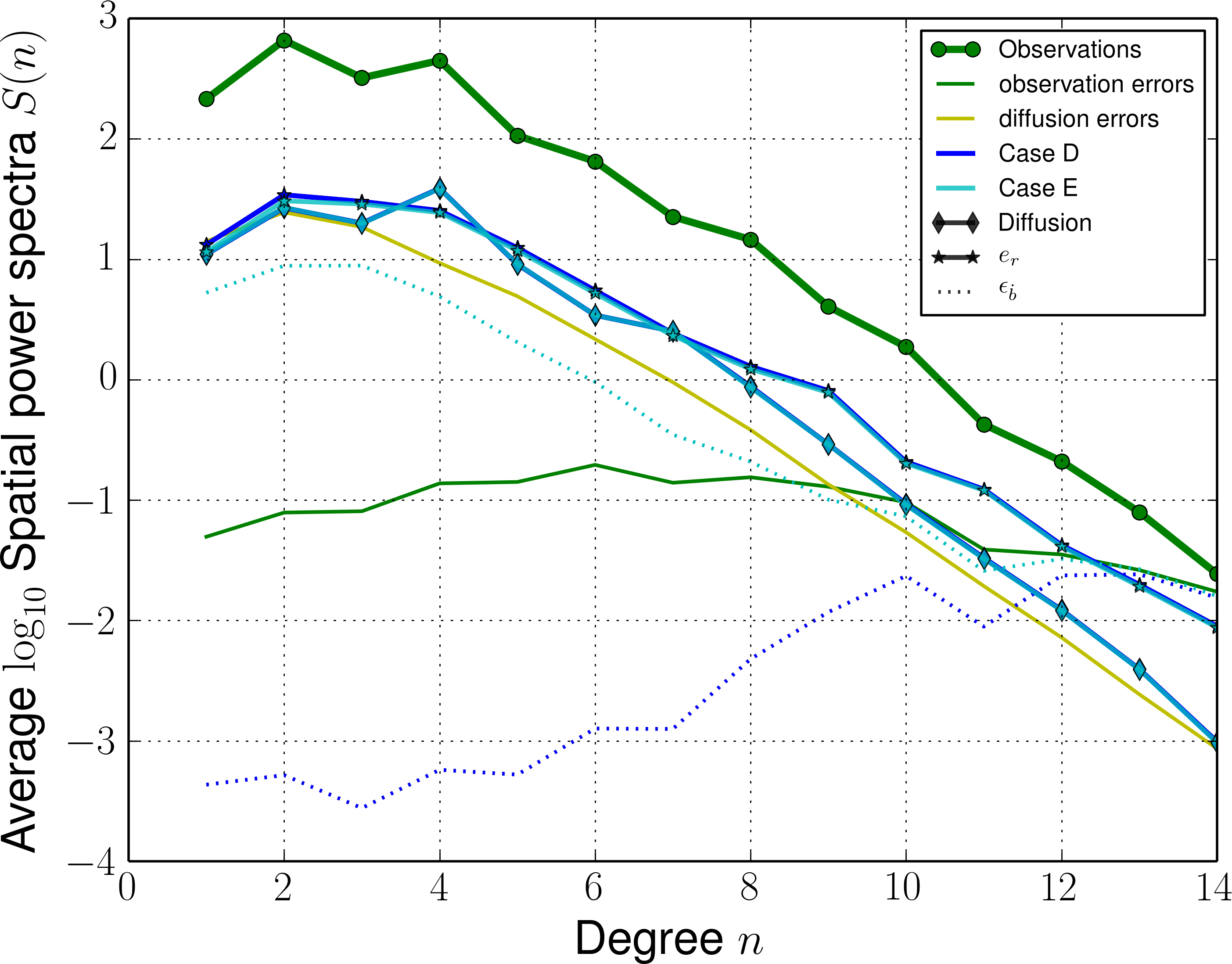}
	\caption{Time average SV spatial power spectra: 
spectrum $\displaystyle \left<{\cal R}_*\right>(n)$ for the reference SV trajectory (green circled thick line), for observation errors (thin green line), and for our estimate of the errors on diffusion (thin yellow) obtained from the diagonal elements of ${\sf P}_{dd}^a$. 
We show in blue (cases D) and cyan (case E) the spectra, at the analysis step, for the contributions from diffusion $\displaystyle \left<{\cal D}\right>(n)$ (diamonds), from subgrid errors $\displaystyle \left<{\cal E}\right>(n)$ (stars), and for the dispersion within the ensemble of SV predictions (dotted lines) . 
}
	\label{fig:Synt_13_SV}
\end{figure*}

We first focus on the impact of subgrid errors.
If no significant differences on the average SV prediction and the SV forecast dispersion is observed between cases A and B, ignoring $e_r$ in the model state generates a significant bias between the analysed diffusion and the diffusion of the reference trajectory. 
This is illustrated with Figure \ref{fig:Synt_13_dg3} (top left and bottom left), where we show time series for the several SV contributions to $\dot{h}_1^1$ -- a coefficient representative of the typical behaviour observed in synthetic series, and the dynamics of which is rich enough to make clear the distinction between SV sources.  
Indeed, the analysed SV contribution from diffusion in case B shows, for coefficients of all degrees, important offsets at some epochs (e.g. from 1980 onwards on $\dot{h}_1^1$ series). 
On the contrary, we manage to recover a significant amount of the reference diffusion when including $e_r$ in the core state, with a dispersion that most of the time encompasses the reference diffusion. 
We thus conclude that accounting for $e_r$ is mandatory to obtain an unbiased estimate of the a posteriori diffusion PDF. 
In cases A and D, where both $e_r$ and diffusion are analysed, we obtain a similar performance on the diffusion estimation (see Table \ref{tab:cases}). 

\begin{figure*}
\centering
	\includegraphics[width=1.0\linewidth]{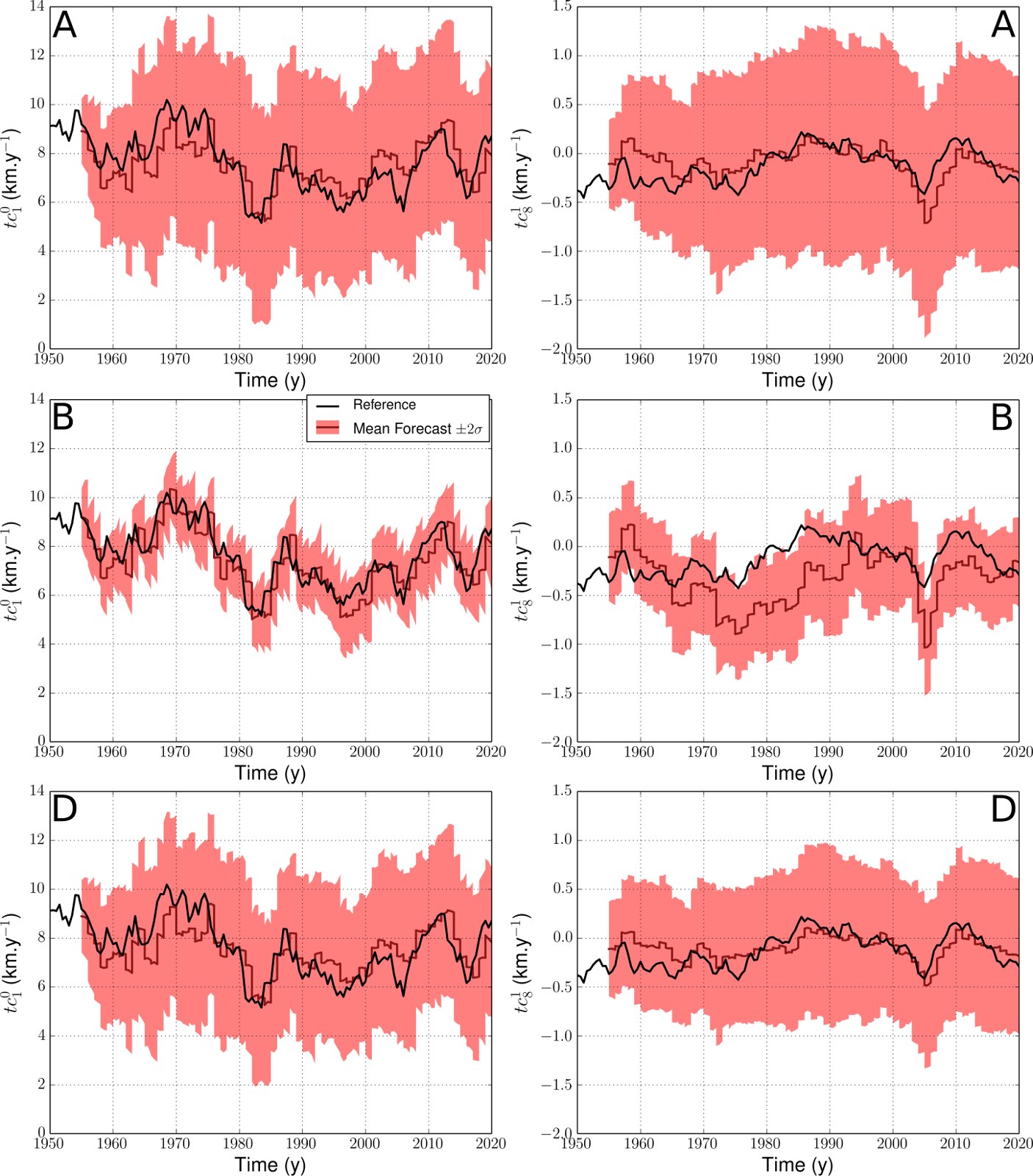}
	\caption{Core flow coefficients series from synthetic experiments for $t_{1}^{0c}$ (left) and $t_{8}^{1c}$ (right):
comparison of $N_m = 50$ reanalyses (ensemble mean in dark red, $\pm 2 \sigma$ in light red) with the reference trajectory (black), in cases A (top), B (middle) and D (bottom).}
	\label{fig:Synt_13_u1-65}
\end{figure*}

\begin{figure*}
\centering
	\includegraphics[width=.7\linewidth]{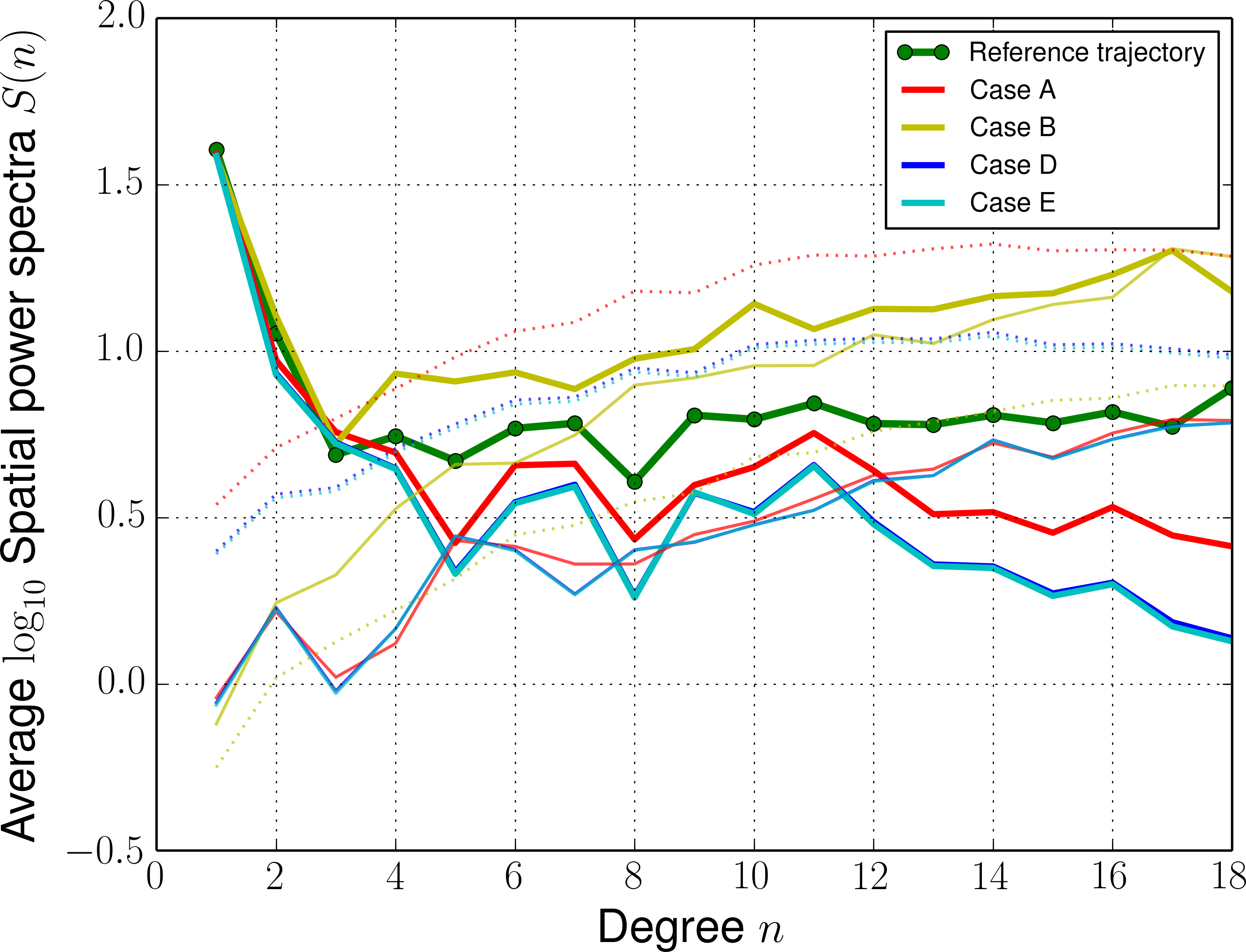}
	\caption{Core flow time average power spectra $\displaystyle \left<{\cal S}\right>(n)$ in cases A (red), B (yellow), D (blue) and E (cyan), for the ensemble average $\hat{\bf u}$ (thick lines), the analysis error ${\bm \delta}_u$ (thin lines) and the dispersion within the ensemble ${\bm \epsilon}_u$ (dotted lines). 
The green circled line shows $\displaystyle \left<{\cal S}\right>(n)$ for the reference trajectory ${\bf u}^*$.
Blue and cyan spectra almost superimpose.}
	\label{fig:Synt_13_SU}
\end{figure*}

Figure~\ref{fig:Synt_13_SV} presents in case D the power spectra of the several contributions to the SV. 
It confirms that the power stored into the analysed diffusion is about $10\%$ that of the observed SV at all length-scales. 
The magnitude of the subgrid errors, similar to that of diffusion at low degrees, appears slightly larger towards small length-scales ($n\ge9$). 
In Figure \ref{fig:Synt_13_u1-65} (upper and middle rows) we show examples of flow coefficients time series, accounting or not for $e_r$. 
Ignoring subgrid errors, we find a significant bias between the reference and analysed flows for all but the largest length-scale coefficients: the reference flow trajectory lays outside the a posteriori distribution provided by the ensemble spread. 
Accounting for $e_r$, this inconsistency is cancelled.
The bottom row of Figure~\ref{fig:Synt_13_u1-65} shows example of flow estimates in case D: if the spread within the ensemble of analyses has been reduced, the ensemble of solutions nevertheless encompasses the reference trajectory at all periods, showing that scaling covariance matrices has helped to better target the reference trajectory.
The good fit to SV changes with a biased analysis in case B arises at the expense of a strong aliasing: the analysed core flow shows too large a power spectrum from spherical harmonic degree $n\ge4$, as illustrated in Figure \ref{fig:Synt_13_SU}. 
This drawback disappears as $e_r$ is reinstated in the model state (case A). 
By scaling matrices (case D), we obtain a flow solution presenting an even lower average spectrum, without increasing the analysis error (i. e. a simpler solution as close to the reference trajectory).

\begin{figure*}
\centering
	\includegraphics[width=.7\linewidth]{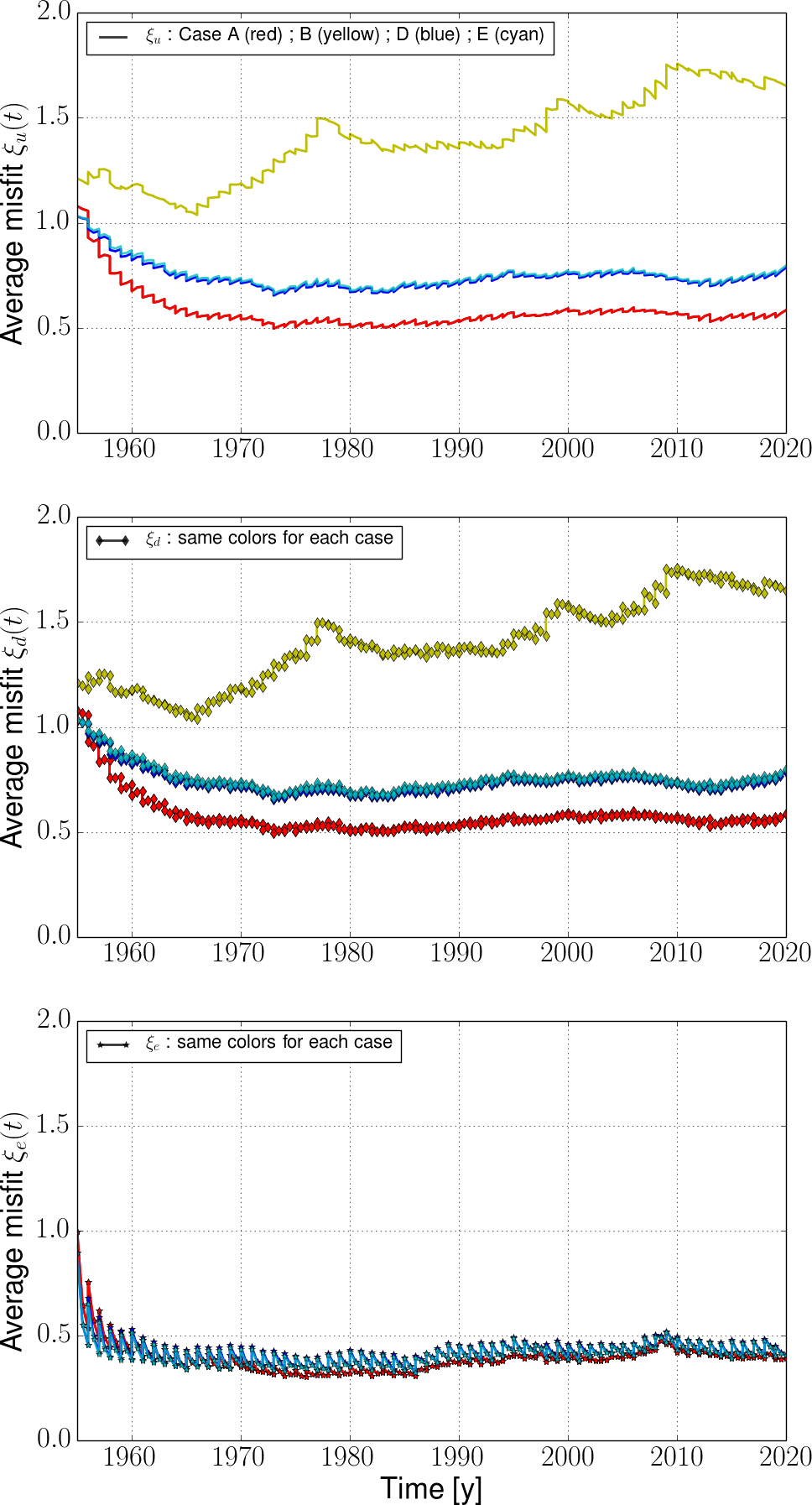}
	\caption{Time evolution of the misfits $\xi_u$ (top), $\xi_d$ (middle) and $\xi_e$ (bottom), given in equation (\ref{eq: misfit_xi}), in cases A (red), B (yellow), D (blue) and E (cyan).
}
	\label{fig:Synt_13_xi}
\end{figure*}

As mentionned in \S\ref{sec: EnKF}, one may wonder whether using scaled matrices would not lead to under-estimate a posteriori uncertainties.
This is actually not the case, as illustrated in Figure \ref{fig:Synt_13_SU} where, for cases A and D, the spectrum for the spread within the ensemble of flow solutions is larger than the spectrum for the bias between ${\bf u}^*$ and $\hat{\bf u}$. 
The same spectra in case B clearly lead to discard this configuration. 
The dispersion seems slightly less over-estimated in case D than in case A. 
This observation is confirmed with the diagnostics $\xi_{u,e,d}$ of equation (\ref{eq: misfit_xi}), shown in Figure \ref{fig:Synt_13_xi} as a function of time, for cases A, B and D:
in case A (resp. D) we over-estimate by a factor about 1.8 (resp. 1.4) the uncertainties on the flow and on diffusion (i.e. the posterior dispersion is a bit conservative), while it is strongly under-estimated in case B. 
We also over-estimate the uncertainties on subgrid errors (by a factor about 2) in both cases A and D. 
Note a warm-up period of about 5 to 10 years before the algorithm reaches approximately steady misfit values.
If both cases D and A show similar scores in Table \ref{tab:cases} for the flow and diffusion, the diagnostics $\xi_{u,d}$ tend to favour case D. 

\begin{figure*}
\centering
	\includegraphics[width=1.0\linewidth]{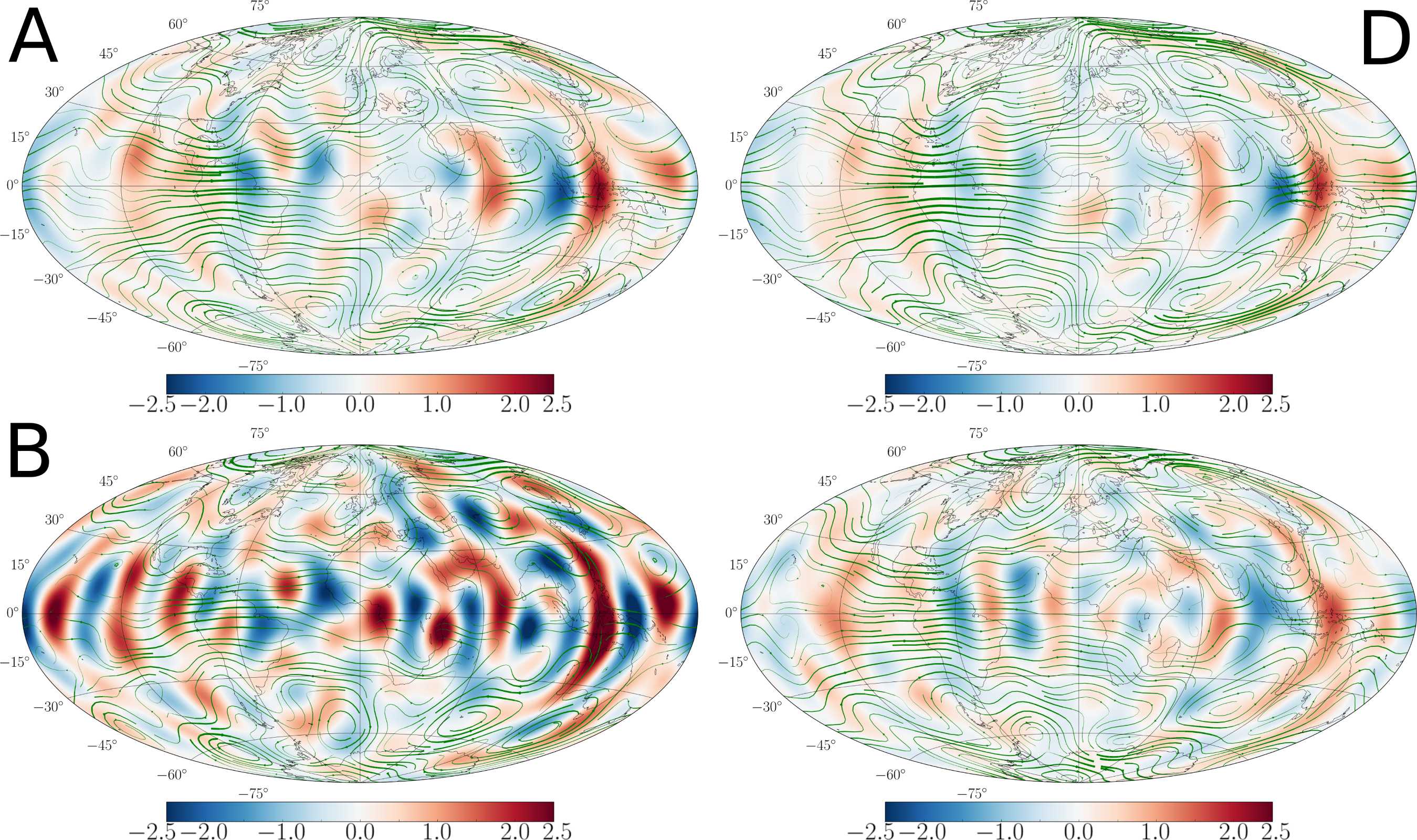}
	\caption{Maps of the horizontal divergence of the flow (red-blue color-scale, in century$^{-1}$) and of the flow (green streamfunction) at the CMB, for the average analysis in 2015, in cases A (top left), B (bottom left), D (top right) and for the reference trajectory (bottom right). The thicker the streamfunction, the stronger the velocity norm (rms velocity over the CMB: 13.28 km/yr).}
	\label{fig:Synt_13_Map_uDivhu}
\end{figure*}

We observe also in the spatial domain the bias observed in the spectral domain, as illustrated with the snapshot surface flow maps in Figure \ref{fig:Synt_13_Map_uDivhu}: 
cases A and D (including $e_r$) are much closer to the reference trajectory than case B (no $e_r$).
The strong aliasing in case B is obvious on the map of the horizontal divergence. To a lesser extent, case A also shows a larger amount of meanders than the simplest case D.
The strong bias obtained for the average model as $e_r$ is ignored is confirmed by normalized misfit values larger than unity for both diffusion and core motions (see Table \ref{tab:cases}). 
On the contrary, the three cases  A, C and D accounting for $e_r$ show far less biases for both observed and unobserved quantities: the relative error on core motions for degrees $n\le 8$ decreases to about 30\%. 
However, since the power in core flows is larger towards long periods, the misfits and spectra discussed so far are  dominated by the time-average state, and give little information about the time changes of the core state.

\begin{figure*}
\centering
	\includegraphics[width=1.0\linewidth]{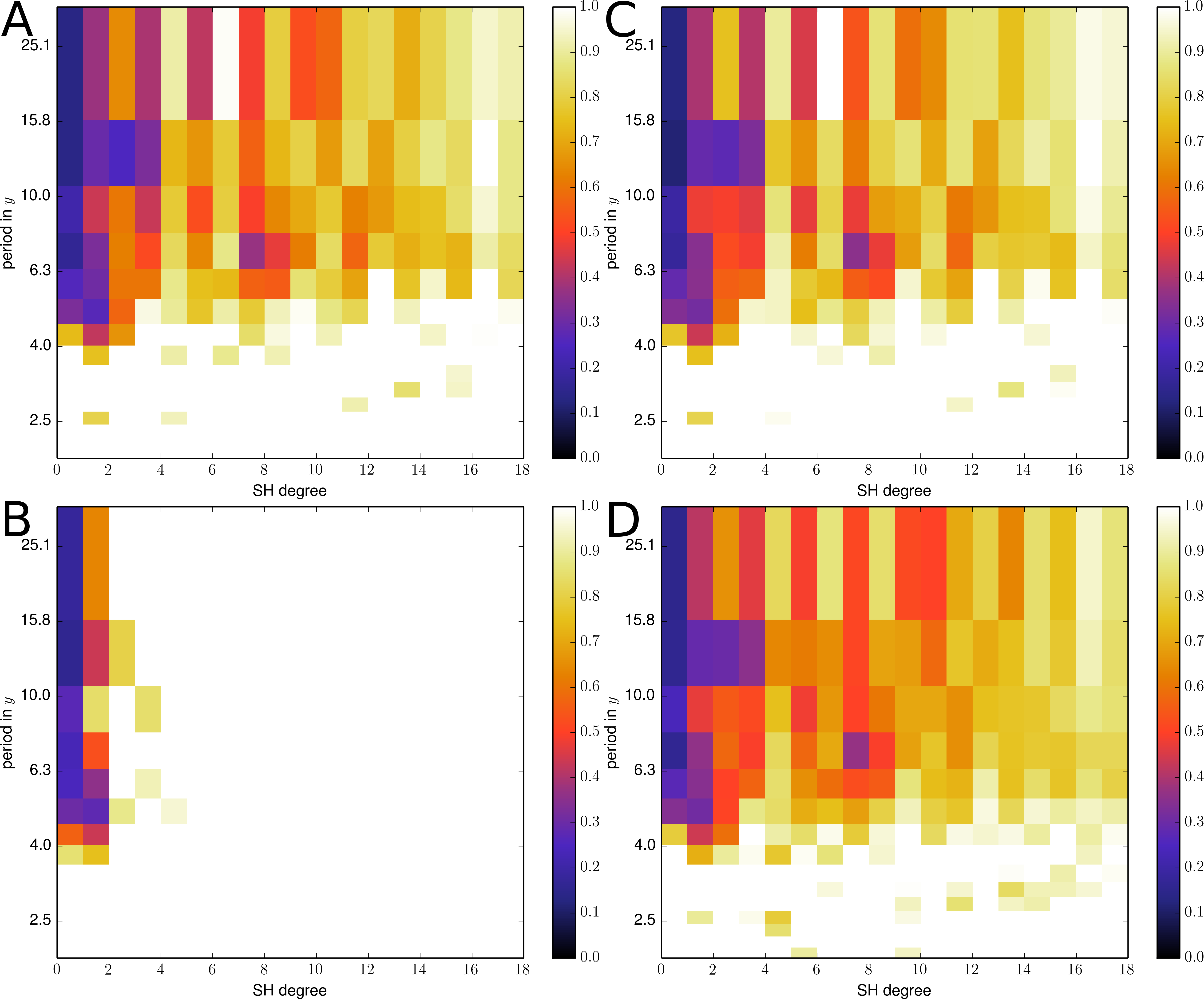}
	\caption{Resolution function ${\cal C}$ as a function of spherical harmonic degree and period, in cases A (top left), B (bottom left), C (top right) and D (bottom right). Black (resp. white) corresponds to 0\% (resp. 100\%) difference between the reference and analysed trajectories.}
	\label{fig:Synt_13_FU}
\end{figure*}

We now investigate more closely the core flow resolution as a function of wave number and period, and present in Figure \ref{fig:Synt_13_FU}, for the four cases, the ratio ${\cal C}(n,f)$ defined by equation (\ref{eq: chart (f,n)}). 
The comparison proposed in Figure \ref{fig:Synt_13_FU} (top left and bottom left) clearly stresses that ignoring $e_r$ (case B), almost no information on flow fluctuations is retrieved from degree $n\ge3$, while a decent amount of information is obtained up to degree $n\simeq 10$ for the lowermost frequencies in case A. 
We also visualize with Figure \ref{fig:Synt_13_FU} (top right) that ignoring diffusion but accounting for $e_r$ (case C) generates a much less severe mismatch than ignoring $e_r$ but including diffusion (the worst case B). 
This confirms that a significant part of the flow may be retrieved under the frozen flux approximation even when this assumption is not exact \citep[and see the snapshot core flow inversions from dynamo simulations by][]{rau2000core}. 
Still, since improving our knowledge of the flow indirectly improves our estimate of diffusion, we obtain a slightly better reanalysis in case A than in case C: we conclude that if it is mandatory to include $e_r$ in the core state, it is also worth accounting for diffusion in our algorithm. 

\begin{figure*}
\centering
	\includegraphics[width=1.0\linewidth]{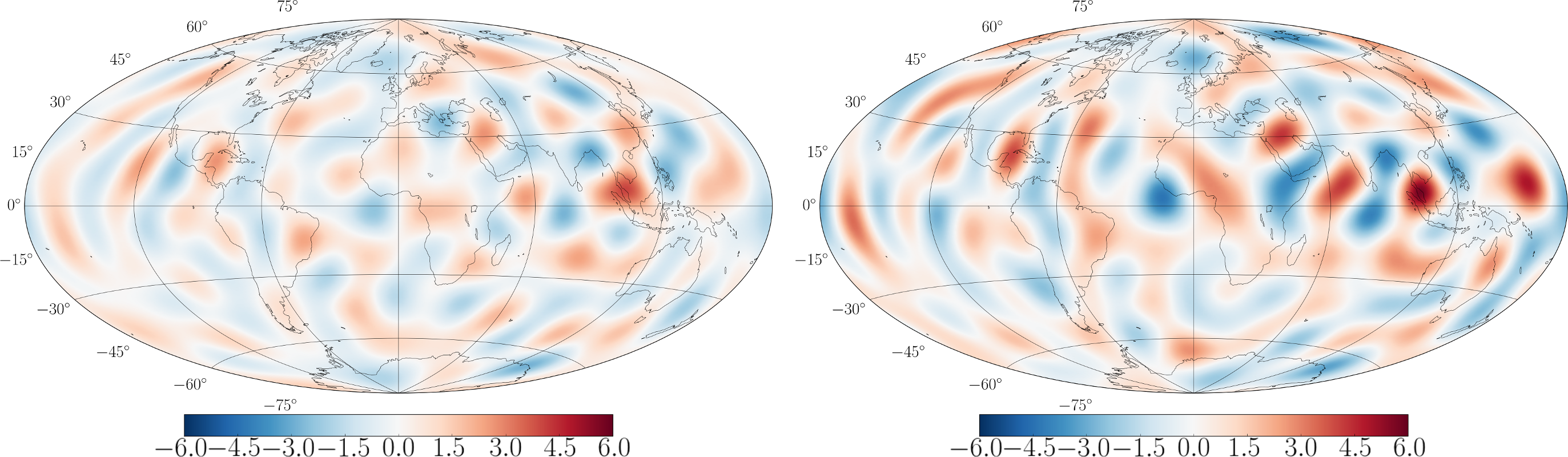}	
	\caption{Map of diffusion (nT/y)  at the CMB from our analysed state in 2015: reference state (right) and ensemble average analysis in case D (left).}
	\label{fig:Synt_13_Map_Diff}
\end{figure*}

We now specifically focus on case D with both $e_r$ and diffusion, but scaling the model covariance matrices according to the stochastic prior dispersion (see \S\ref{sec: EnKF}).
While the scaling (\ref{eq: scaling}) only marginally improves diagnostics dominated by long periods (see Figure~\ref{fig:Synt_13_SU} and Table~\ref{tab:cases}), Figure \ref{fig:Synt_13_FU} (bottom right) shows that it allows us to slightly better recover rapidly changing flow patterns, especially towards small length-scales. 
We witness here that allowing at each analysis step for a too large innovation (the prior constraint on the model increment in cases A to C is weaker than in case D), we lose some constraints on the transient motions. 
For those reasons, even if no significant improvement is seen for slow core flow changes, we are inclined to favour case D (which also provides simpler solutions and misfits $\xi_{u,d}$ closer to one).
We compare in Figure \ref{fig:Synt_13_Map_Diff}, for our preferred case D, the spatial distribution of the contribution from diffusion to the SV at the CMB. 
We overall find the correct amplitude (of the order of $\pm 5$ nT/yr), and are able to localize some of the main diffusion patches, as for instance in the Eastern hemisphere between $\pm 40^{\circ}$ latitude. 
The largest patterns appear in the equatorial area. 
These are found to correlate well with the main up/down-wellings at the CMB (compare the maps of diffusion and $\nabla_h\cdot({\bm u}_H)$ in Figure \ref{fig:Synt_13_Map_uDivhu}). 

We finally compare case D to the last configuration E, where in addition errors on the analysis of diffusion are accounted for. 
Surprisingly, we see very little changes concerning both the scores of Table \ref{tab:cases},  the diagnostics in Figure \ref{fig:Synt_13_xi}, resolution charts ${\cal C}(n,f)$ (not shown) or the flow spectra (Figure \ref{fig:Synt_13_SU}). 
The latter almost superimpose in the two cases not only for the ensemble average flow, but also for the flow dispersion and the average analysis error. 
Interestingly, the ensemble average diffusion and subgrid errors (as well as their associated dispersion) are also very similar in the two cases (see figure \ref{fig:Synt_13_dg3}). 
The main difference concerns the SV prediction: if these are in average similar in the two cases, a much larger dispersion is found in case E than in case D (see Figure \ref{fig:Synt_13_dg3}). 
This behaviour derives from the much looser constraint imposed in case E on the fit to SV data (through ${\sf R}_{yy}$), and is characterized by enhanced model prediction errors in case E (see Figure \ref{fig:Synt_13_SV}). 

\subsection{Geophysical application}
\label{sec: geophys}

We now apply our algorithm to an ensemble of realizations of the geomagnetic field model COV-OBS.x1, from $t_s=1950$ to $t_e=2020$. 
The model prior is the same as that used for the synthetic experiment, i.e. the configuration of case D  (unless specified otherwise) with $\tau_u=30$ yrs and $\Delta t^a=1$ yr. 
As in the synthetic experiments, analysed flow and diffusion are very similar in cases E and D except for SV predictions, and we only show results for the latter configuration. 
Performing inversions with instead $\tau_u=100$ years, i.e.  with a pre-factor $\alpha_u$ of 0.020 instead of 0.067 in equation (\ref{eq: scaling}), only minor changes are observed on the ensemble average solution. 

\subsubsection{Contributions to the secular variation}
\label{sec: SV contributions}

\begin{figure*}
\centering
	\includegraphics[width=1.0\linewidth]{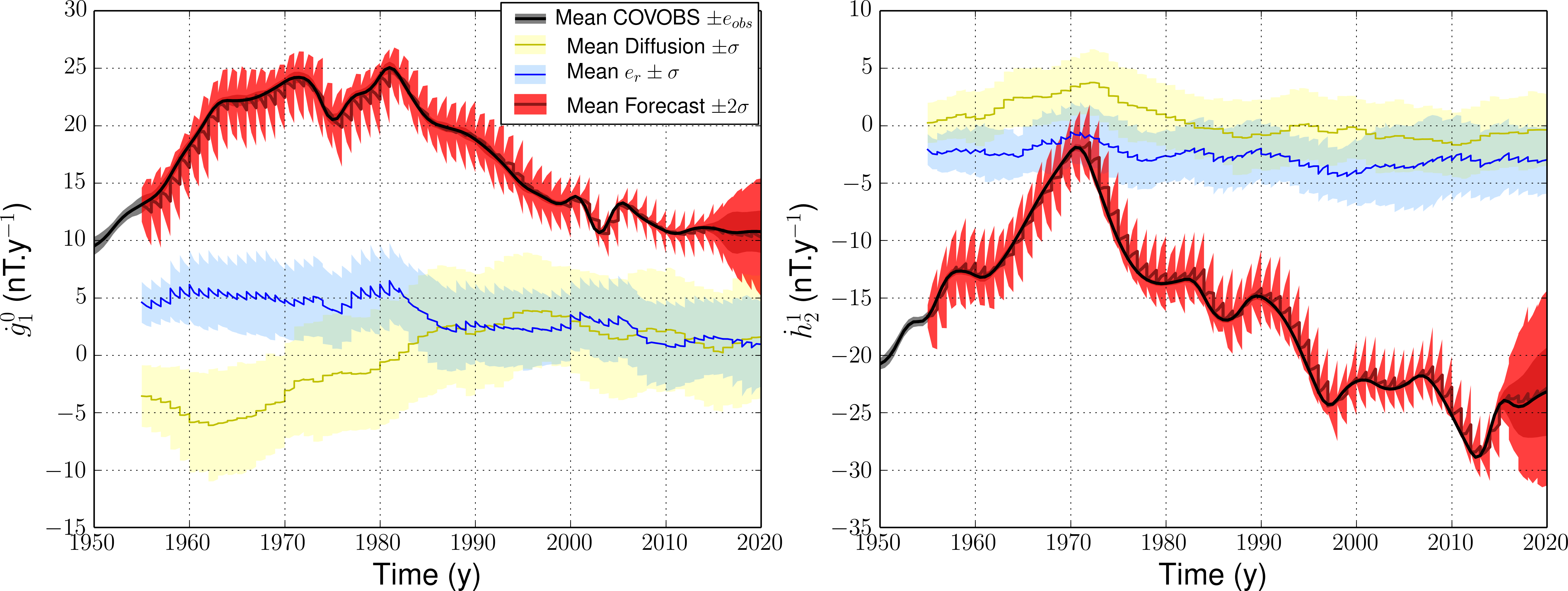}
\centerline{
	\includegraphics[width=.5\linewidth]{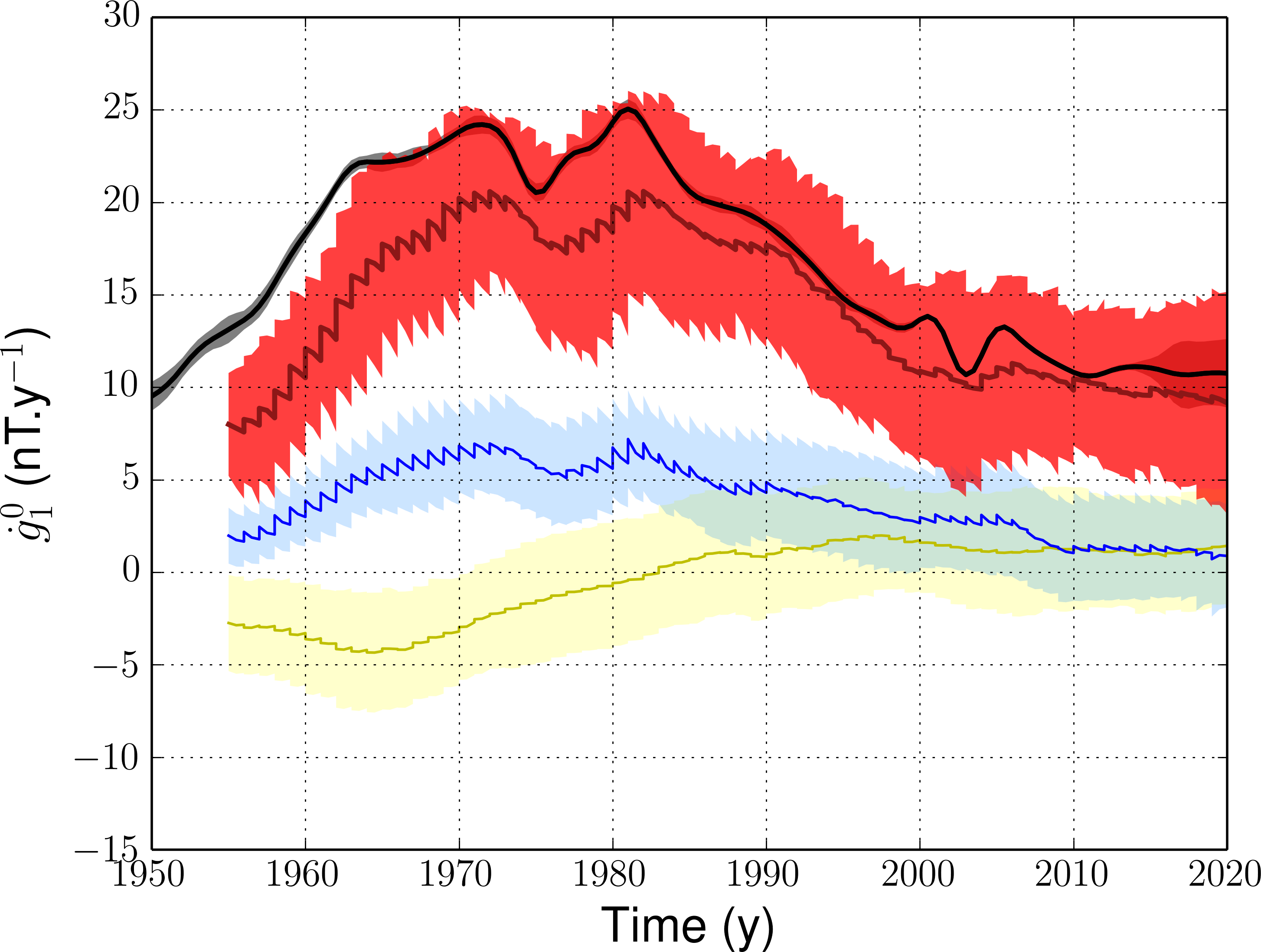}
	\includegraphics[width=.5\linewidth]{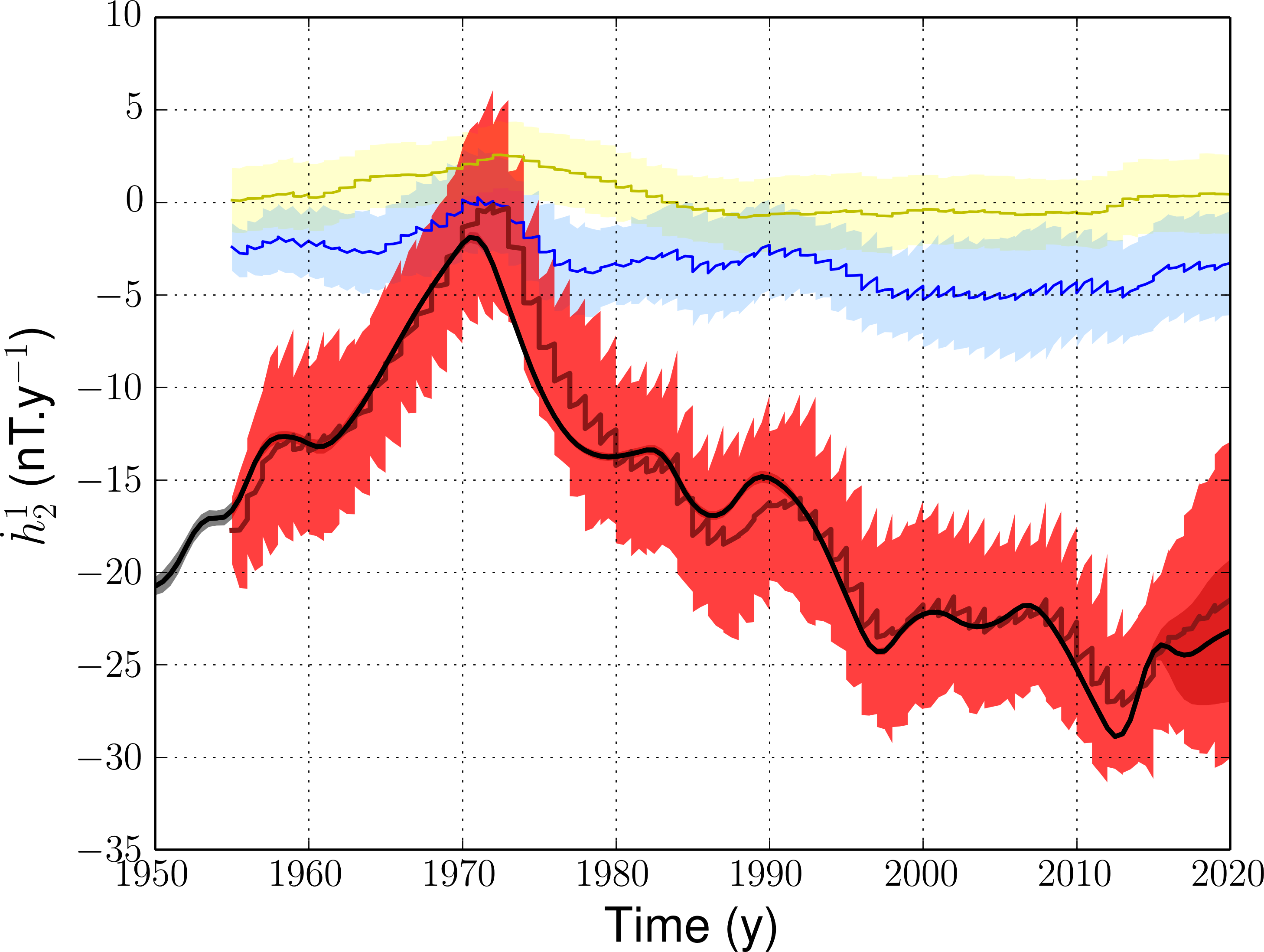}}

	\caption{Series of SV coefficients $\dot{g}_{1}^{0}$ (left) and $\dot{h}_{2}^{1}$ (right): for the COV-OBS.x1 model (average in black, $\pm\sigma$ in grey), and $N_m$ forecasts from our assimilation algorithm (average in dark red, $\pm 2 \sigma$ in red), in the configurations D (top) and E (bottom). 
In blue (resp. yellow) are shown the estimated contributions from $e_r$ (resp. from diffusion).}
	\label{fig:COV_14_dg1-6}
\end{figure*}

During the whole studied timespan, the dispersion within the ensemble of SV forecasts is large enough to include the observed SV changes within $\pm 2 \sigma$, even when jerks occur during the most accurate satellite era (see figure~\ref{fig:COV_14_dg1-6}). 
Our algorithm thus provides a coherent estimate of the PDF for the SV coefficients in this geophysical context.
Subgrid errors and diffusion both represent about $20\%$ of the total dipole decay, and potentially contain a non-zero average contribution. 
The same observations holds for a non-dipole SV coefficient such as  $\dot{h}_2^1$, shown in figure~\ref{fig:COV_14_dg1-6} (right). 
Note that COV-OBS.x1 from 2015 onwards is the result of a prediction, built on magnetic records prior to 2014.6 and on the time cross-covariances of the magnetic model prior \citep{gillet2015stochastic}. 
For those reasons, observations errors in our study drastically increase after 2015, leading to the widening of the $\pm 2\sigma$ values in Figure \ref{fig:COV_14_dg1-6}.

As in the synthetic experiments, the spread in SV predictions is larger in case E (figure~\ref{fig:COV_14_dg1-6}, bottom) than in case D. 
We note a shift towards zero of the average axial dipole decay, larger as $\dot{g}_1^0$ reaches large values prior to 1985. 
Still the dispersion in case E encompasses the observed SV, except during the warm-up phase.
It is worth notice that diffusion and subgrid errors show rather similar average trajectories in the two cases E and D, with comparable dispersion. 

Our ensemble of forecasts tends, in average, to drive the system towards low SV values. 
This observation is particularly clear as the recorded SV reaches large values, generating the saw-tooth pattern on $\dot{g}_{1}^{0}$ prior to 1980, and during phases where $| \dot{h}_2^1|$ increases (see Figure \ref{fig:COV_14_dg1-6}).
It is to be expected with the kind of stochastic model we employ, where the most likely flow forecast decays exponentially towards the background flow $\hat{\bm u}_H$ in a time $\tau_u$, driving the average SV forecast naturally to lower values. 
As such, our average model is not designed to present a predictive power.
We associate the better predictions for the dipole decay during the satellite era to the lower value of $\dot{g}_{1}^{0}$ at that epoch, and to an observed SV decreasing in a similar manner to the AR-1 model.  

Our re-analysis confirms that the dipole decay is primarily driven by advection, as suggested in \cite{finlay2016gyre}.
Nonetheless, we find a non-zero negative contribution from diffusion to the dipole decay before 1980 (down to $-6$ nT.y$^{-1}$ in the early 1960s).
This observation contrasts with the previous estimate by \citeauthor{finlay2016gyre}, who found a diffusion contribution almost stationary at about $+5$ nT/yr.
The difference reflects the impact of flow motions on the analysis of diffusion (see \S\ref{sec: diffusion}).
For the most recent and best documented epochs since 2000, where $\dot{g}_1^0$ reaches lower values (from 10 to 15 nT/yr), we still find that diffusion is not the major source of the dipole decay.

\subsubsection{Magnetic diffusion and westward gyre}

\begin{figure*}
\centering
	\includegraphics[width=1.0\linewidth]{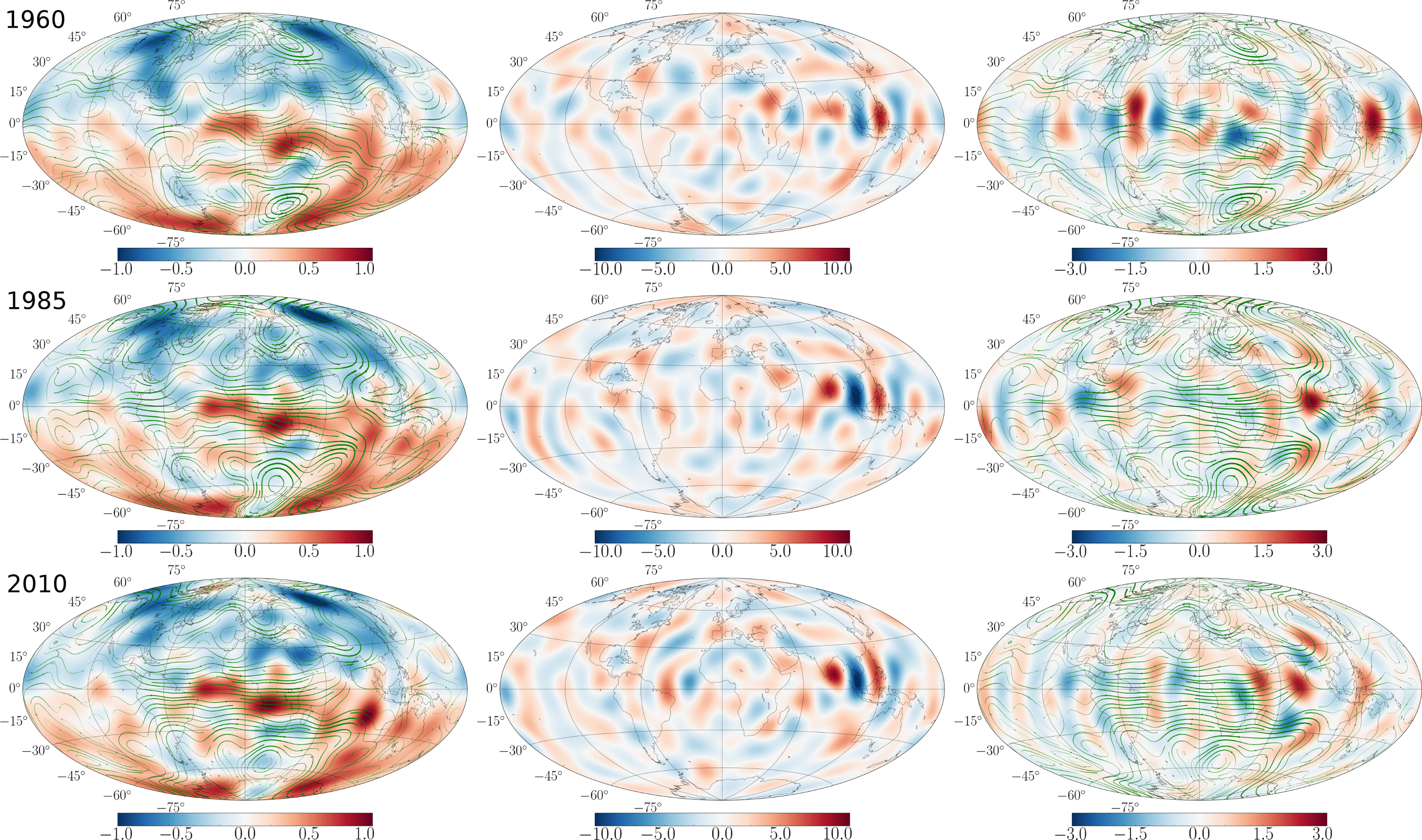}
	\caption{
	 Ensemble average re-analyses at epochs $1960,1985$ and $2010$ (from top to bottom) at the CMB from COV-OBS.x1.
Left: maps of the main radial field $B_r$ (red-blue color-scale, in $\mu$T) and of the flow (green streamfunctions). 
Middle: maps of the contribution from diffusion to the SV (in nT/y). 
Right: maps of the horizontal divergence of the flow (red-blue color-scale, in century$^{-1}$) and of the flow (green streamfunctions).
}
	\label{fig:COV_14_Maps}
\end{figure*}

In the spatial domain (see the middle column of figure \ref{fig:COV_14_Maps}), our analysis of diffusion shows localized patches reaching up to $\pm 12$ nT/yr, as for instance below Indonesia.
Again, these are in relation with up/down-wellings (figure \ref{fig:COV_14_Maps}, right column) that primarily shows up in the equatorial area. 
This link is not systematic though, because diffusion is not enslaved only to the flow: it  also depends locally on the magnetic field morphology. 
Indeed, the large up-welling to the North-East of Brazil in 1960 is associated with little diffusion.
The link between up/down-wellings and surface diffusion was suggested by \cite{amit2008accounting}, through the poloidal flow component carried by columnar structures.
However, we do not retrieve the prominent diffusion feature that they highlight below the Pacific.  
Here, we associate the localized diffusion patterns in the equatorial belt with the eccentric westward gyre put forward by \cite{pais2008quasi}.
As \cite{aubert2013flow} and \cite{gillet2015planetary} before us (under respectively a dynamo norm and a QG constraint) we retrieved here this planetary-scale structure in our re-analysis (right column of Figure \ref{fig:COV_14_Maps}). 
We find up/down-welling and the largest signatures of diffusion where the gyre reaches the equatorial area. 
Although influenced by the primarily equatorial symmetric CE dynamo prior, our solution displays in this area a velocity field that crosses the equator, violating locally the QG assumption, in agreement with the conclusions of \cite{baerenzung2016flow}. 

Interestingly, our estimate of diffusion also differs from that of \cite{chulliat2010observation}. 
From the analysis of satellite field models, they found below the South Atlantic ocean violations of topological constraints derived from the assumption of an infinitely conducting outer core \citep[namely changes of the magnetic flux passing through areas delimited by null-flux curves, see for instance][]{jackson1996kelvin}, which they interpret as the signature of diffusion. 
Our solutions do not show particularly large diffusion in conjunction with the South Atlantic Anomaly (see Figure \ref{fig:COV_14_Maps}, left and middle columns). 
We associate the difference with the findings of \citeauthor{chulliat2010observation} to the role played by subgrid processes: they blur our image of null-flux curves, which soften the topological constraints \citep[cf][]{gillet2009ensemble}.
Alternatively, we see a possible correlation between the localized equatorial patches of diffusion and the rapid changes in the secular acceleration \citep[$\partial^2 B_r/\partial t^2$, see][]{chulliat2014geomagnetic,chulliat2015fast}, through flow perturbations around the westward gyre \citep{finlay2016recent}. 
Indeed strong secular acceleration patterns are found under Indonesia and Central to South America, the location where we also isolate the strongest diffusion and up/down-welling features. 

Figure~\ref{fig:COV_14_Maps} shows that the westward gyre is present since 1960, suggesting a temporal stability of the largest flow features (the rms velocity over the CMB in $1960, 1985$ and $2010$ are respectively 13.9, 14.0 and 12.3 km/yr).
However, towards the most recent epochs it strengthens below South America and the Atlantic ocean, at the same time the large up-welling present below NE Brazil around 1960 vanishes. 
We also notice the occurrence of secondary circulations with decadal time scales, such as the vortices below $30^{\circ}$ latitude in the Eastern Pacific hemisphere and those centred around $\pm30^{\circ}$ latitude in the western Pacific, which are present in 1985 but have almost disappeared in 2010. 
The westward gyre also appears as a complex structure, with modulation of its meanders throughout the studied era.
Even though our ensemble average solution does not capture the fastest changes in the core trajectory at small length-scales (cf Figure \ref{fig:Synt_13_FU}), maps shown in Figure \ref{fig:COV_14_Maps} suggest nevertheless that some time-dependent meso-scale eddies seem to be robust \citep[see][]{gillet2015planetary,amit2013differences}.

\subsubsection{Length-of-day predictions}

\begin{figure*}
\centering
	\includegraphics[width=.7\linewidth]{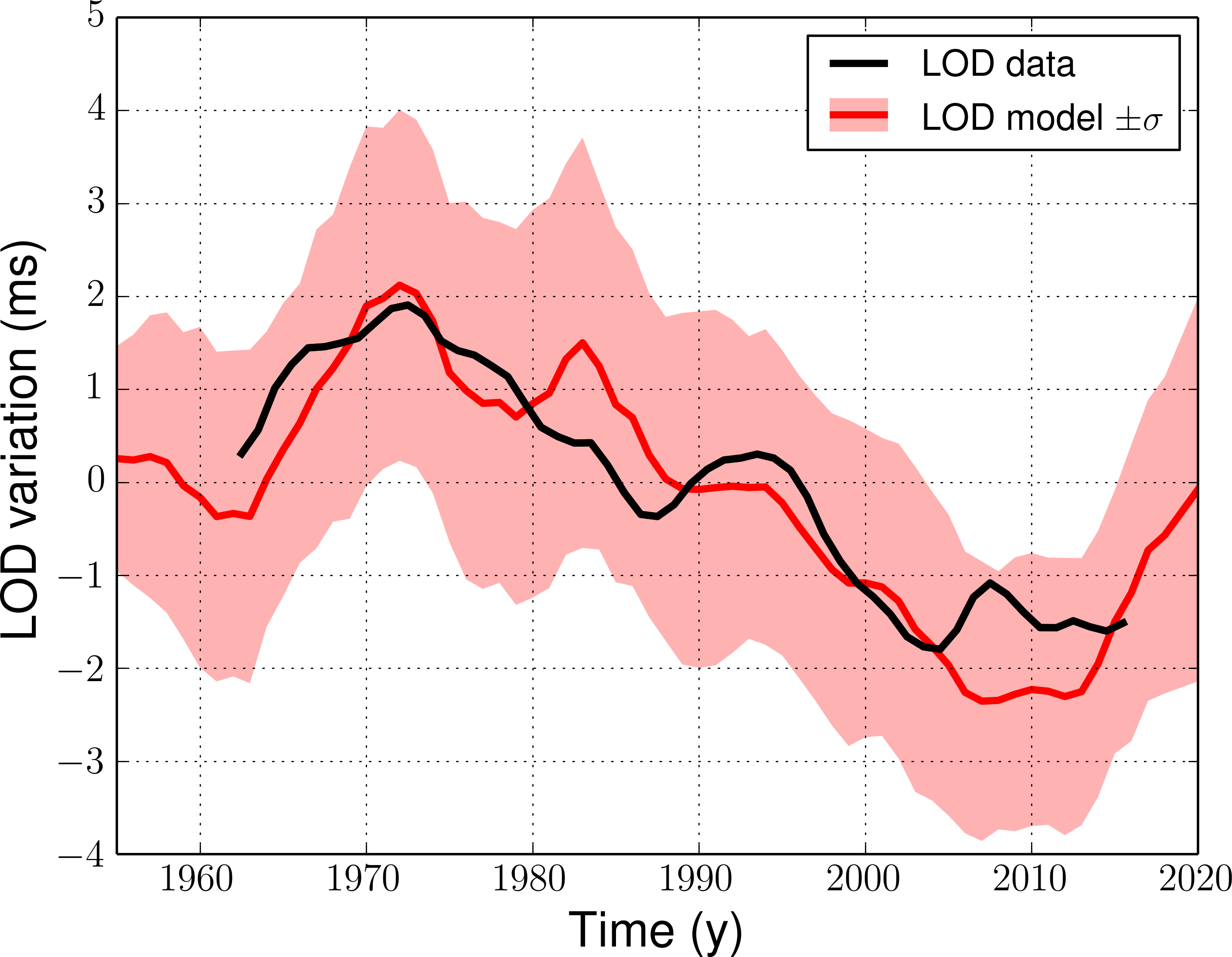}
	\caption{Length of Day (LOD) variations predicted from our ensemble of re-analyses (average in red, $\pm\sigma$ in red shaded area), compared with the geodesic data (black). Notice that we only plot here the LOD of the re-analysis (and not of forecasts).}
	\label{fig:COV_14_LOD}
\end{figure*}

We now confront the result of our re-analysis to an independent geophysical observation, namely changes in the length-of-day (LOD). 
LOD data are here computed from annual means of angular momentum series provided by the IERS \citep[the C04 series, see][]{bizouard2009combined} cleaned for solid tides (the IERS 2000 model)
and for atmospheric predictions from the NCEP/NCAR re-analysis \citep[see][and references therein]{zhou2006revised}.
A 1.4 ms/cy trend has been removed, corresponding to the observed LOD trend over the past centuries \citep{stephenson1984long}.
LOD predictions from our ensemble of flow models are computed using equation (101) of \cite{Jault:2015kq}, which accounts for the effect of compressibility on the radial density profile -- though very little difference is found with the original formula by \cite{jault1988westward}.

Figure~\ref{fig:COV_14_LOD} shows that during the whole studied timespan our model provides a convincing prediction for the decadal LOD changes. 
The recorded geodetic series is captured within the $\pm \sigma$ predictions, and the 1994 local extremum in the LOD is partially caught by the ensemble average re-analysis -- we have knowledge of no flow model capable of entirely predicting this bump, an issue first put forward by \cite{wardinski2005core}.
Focusing during the satellite era, we also note a mismatch between LOD data and our ensemble average prediction, which does not catch the maximum around 2008, contrary to the QG reconstruction by \cite{gillet2015planetary} -- although the 2008 peak still lays within our $\pm \sigma$ envelope.

\subsubsection{Dispersion of the secular variation over 5 yrs intervals}
\label{sec: 5 yrs}

Finally, we address the spread of our predicted SV to geophysical observations in a configuration where analyses are performed every $\Delta t^a = 5$ yrs. 
We consider below three configurations: that of case D (CE cross-covariances, $\tau_u=30$ yrs), a case F where the CE cross-covariances are multiplied by 4 ($\tau_u=30$ yrs), and a case G similar to case D but with $\tau_u=100$ years instead. 
Note that the estimates for diffusion, subgrid errors and the flow at the analysis step are not significantly different in those three cases, meaning the analysed model is relatively robust. 

\begin{figure*}
\centering
	\includegraphics[width=1.0\linewidth]{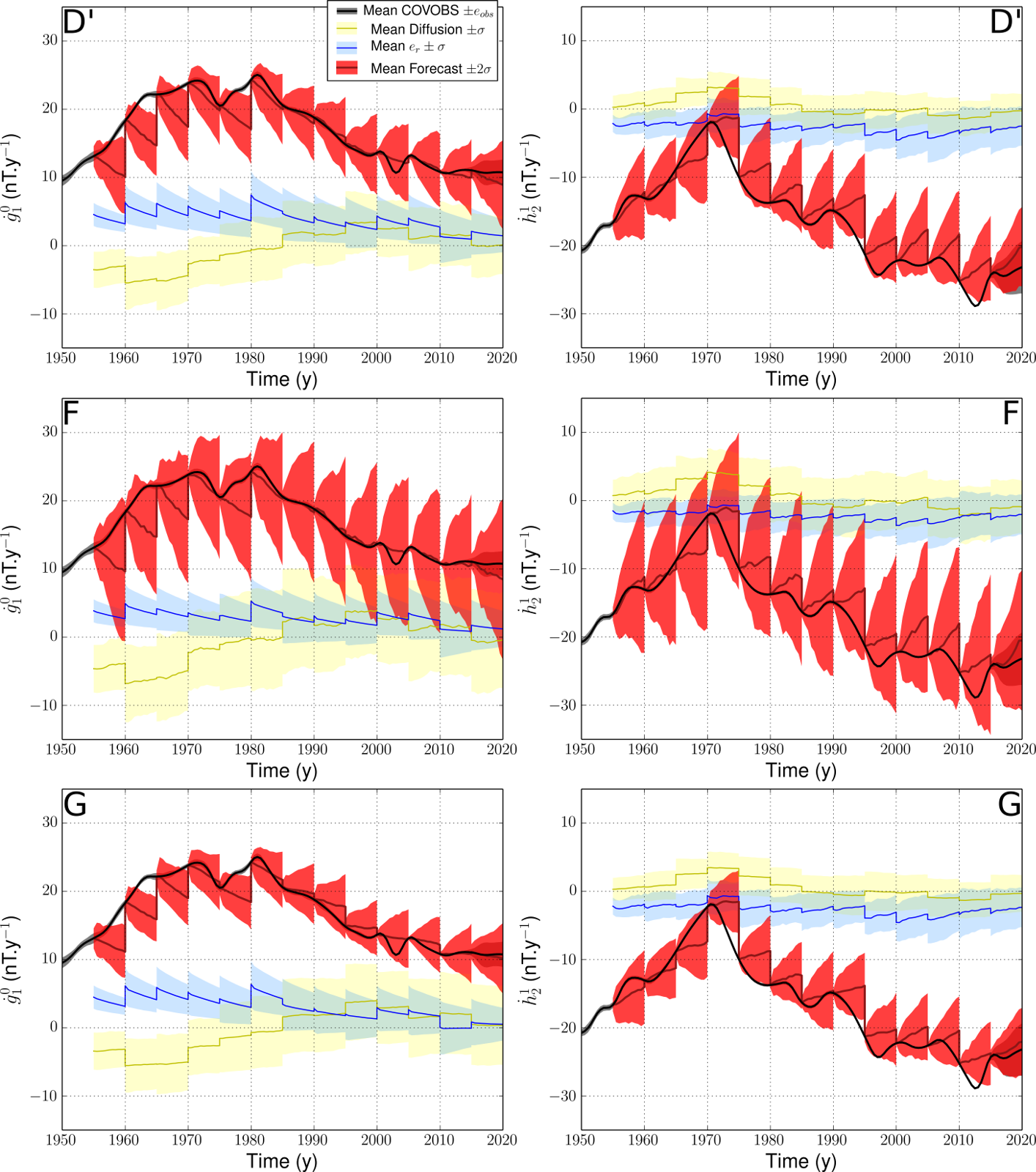}
	\caption{
Series of SV coefficients $\dot{g}_{1}^{0}$ (left) and $\dot{h}_{2}^{1}$ (right) for an analysis window $\Delta t^a = 5$ yrs in case D (top), F (middle) and G (bottom). 
See text for details. 
The legend is the same as in Figure~\ref{fig:COV_14_dg1-6}. }
	\label{fig:COV_14_dg1-6_dta5}
\end{figure*}

SV re-analyses of COV-OBS.x1 data in cases  D, F and G are shown in Figure~\ref{fig:COV_14_dg1-6_dta5}.
In case D, the observed SV is almost always embedded within $\pm2\sigma$ of the 5 yrs SV forecasts for all coefficients but the axial dipole (see Figure~\ref{fig:COV_14_dg1-6_dta5}, top). 
Our model indeed misses the trend towards large $\dot{g}_1^0$ values recorded prior to $1980$ -- in line with the natural behaviour of average SV forecast mentioned in \S\ref{sec: SV contributions}.
This observation suggests three possibilities:
(i) the cross-covariances we use from the CE dynamo do not allow enough freedom,
(ii) the decay towards the background is too fast ($\tau_u$ too small), or
(iii) higher order statistics are needed to mimic the behaviour of the dipole decay in particular, as observed in paleomagnetic records \citep{love2003gaussian} and in numerical simulations \citep[e.g.][]{bouligand2005statistical,fournier2011}. 

We test the first two possibilities with cases respectively F and G. 
The alternative (iii) could be attended using more complex stochastic models \citep[e.g.][]{buffett2013stochastic}. 
By increasing the model covariances (Figure~\ref{fig:COV_14_dg1-6_dta5}, middle), the dispersion within the ensemble of SV predictions is enlarged for all coefficients, and a factor of 2 on the prior model dispersion is enough for the observed dipole decay to lay within $\pm 2 \sigma$.
By increasing $\tau_u$ (Figure~\ref{fig:COV_14_dg1-6_dta5} bottom), the decay of the SV forecast becomes naturally slower, although the dispersion is also reduced, so that case G, as case D, does not always catch the observed $\dot{g}_1^0$ within $\pm 2 \sigma$.

\section{Summary and discussion}
\label{sec: discussion}

\subsection{New insights from our approach}
\label{sec: algo}

Following earlier strategies developed for geomagnetic data assimilation \citep{aubert2015geomagnetic,gillet2015stochastic,whaler2015derivation}, the algorithm we present in this study proposes to mix spatial information from Earth-like geodynamo simulations and a temporal information compatible with the frequency spectrum of geomagnetic series, to re-analyse geomagnetic field models within a stochastic, augmented state Kalman filter.
We have shown from time-dependent synthetic experiments that subgrid errors that arise from interactions between the unresolved magnetic field at small length-scales and core motions must be accounted for.
Indeed, ignoring them leads to strong bias and aliasing in the analysed core state.
By representing sugbrid errors by means of a stochastic equation, we significantly improve our recovery of the time-dependent core state.
Our augmented state algorithm furthermore circumvents the bias encountered for SV predictions by \cite{gillet2015planetary}, who had implemented the stochastic constraints within a weak formalism (i.e. through covariance matrices instead of time-stepped stochastic equations). 

We also propose a new avenue to estimate diffusion at the CMB, from cross-correlations (inferred from geodynamo simulations) between diffusion and both magnetic and velocity surface fields. 
Indeed, diffusion is related to the magnetic field at and below the core surface, and thus is coupled to core motions.
We show from synthetic experiments that a non-negligible amount of diffusion can be recovered. 
Our analysis furthermore suggests that diffusion must carry a high frequency content, through its link with up/down-wellings.
Its amplitude is locally as large as about 10 nT/yr.
Our analysis shows rather different estimations of diffusion in comparison with previous studies: as mentioned in \S\ref{sec: geophys}, we find no crucial signature of diffusion associated with the South Atlantic anomaly, but instead a significant contribution on the equator below Indonesia.  

\subsection{Future evolution of the algorithm}

The tool we derived remains nevertheless imperfect, which calls for future methodological developments. 
Our algorithm indeed does not integrate all the power of the EnKF, essentially in link with our crude estimate of the analysis error cross-covariances (see \S\ref{sec: EnKF}). 
Our attempts at approximating these more closely (e.g. using ${\sf P}_{zz}^f= \alpha {\sf P}_{zz} + \left[{\sf I}-{\sf K}_{zz}{\sf H}\right]{\sf P}_{zz}$, not shown) actually under-perform the simpler representations with frozen matrices.  
This calls for alternatives to localize cross-covariances in the spectral domain \citep{wieczorek2005localized} if one wishes to avoid computing thousands of realizations.    

We have investigated the impact of errors on the analysed diffusion, which should in principle be considered, with a crude estimate of their cross-covariances. 
We found that adding their contribution to the observation error -- through the matrix ${\sf R}_{yy}$ in equation (\ref{eq:EnKF step 2}) -- only marginally modifies the solutions for the flow and diffusion (average and dispersion), while allowing for a larger SV spread at the analysis steps.  
This difference is most probably accommodated by the flexible representation of time-correlated model errors through an augmented state -- which possibly ingest other sources of uncertainties that are not explicitly accounted for. 
Accordingly, the impact on the spread in 5 years (or longer) SV forecasts may appear negligible in comparison with uncertainties associated with our choice of prior information (see \S\ref{sec: 5 yrs}). 

\subsection{An hypothesis testing tool}
\label{sec: benchmark}

However, the estimate of the surface core trajectory (flow and diffusion) will depend on the considered geodynamo model. 
In particular the sensitivity of the analysed diffusion to the chosen dynamo prior calls for further investigations using dynamo simulations run at more extreme parameters \citep[e.g.][]{aubert2017spherical,schaeffer2017geodynamo}.
Our algorithm is by construction flexible: any spatial cross-covariances may be considered.
Indeed, one only needs well-conditioned statistics on $B_r$, ${\bm u}_H$, $e_r$ and $\eta \nabla^2 B_r$ from any forward model to test different hypotheses, such as the amount of quasi-geostrophy, the need for an asymmetric thermal forcing, etc.

Note also that our algorithm allows for possible changes in the forward (time-integrated) stochastic model. 
Alternatives to our simple AR-1 representation may be considered, for both subgrid errors and core motions (cf section \S\ref{sec: 5 yrs}).
Furthermore, our AR-1 model may be used as a zero-state for comparisons with algorithms using deterministic equations. 
One could for instance measure if assimilation tools based on prognostic geodynamo models \citep[e.g.][]{fournier2013ensemble} perform better than the same dynamo spatial statistics embedded in our stochastic algorithm, in either a re-analysis or a forecast framework.
 
\subsection{Towards an operational assimilation tool}

In the perspective of developing operational geomagnetic data assimilation tools, our algorithm may be seen as a first step before ingesting direct magnetic records (from satellites, observatories, etc.), instead of their interpolation through Gauss coefficients as in the present study. 
This may require not only the migration of observations at each analysis epochs \citep[as done with virtual observatories, see][]{mandea2006new}, but also the co-estimation of external sources together with core motions, which calls for further developments. 
Despite the limitations of its predictive power, we can envision with the strategy developed throughout this study to build IGRF candidate models \citep[in particular the SV for the coming 5 years together with its associated uncertainties, see][]{thebault2015international} constrained by core motions. 

\section{Acknowledgements}

NG and OB were partially supported by the French Centre National d'Etudes Spatiales (CNES) for the study of Earth's core dynamics in the context of the Swarm mission of ESA. 
ISTerre is part of Labex OSUG@2020 (ANR10 LABX56), which with the CNES also finance the phd grant of OB. 
Numerical computations were performed at the Froggy platform of the CIMENT infrastructure (https://ciment.ujf-grenoble.fr) supported by the Rh\^one-Alpes region (GRANT CPER07 13 CIRA), the OSUG@2020 Labex (reference ANR10 LABX56) and the Equip@Meso project (referenceANR-10-EQPX-29-01). 
This is IPGP contribution no XXXX. 

\bibliography{artbib}

\bibliographystyle{gji}

\label{lastpage}

\end{document}